# TALOS (Total Automation of LabVIEW Operations for Science): A framework for autonomous control systems for complex experiments

M. Volponi ✉ ; J. Zieliński ; T. Rauschendorfer ; S. Huck ; R. Caravita ; M. Auzins; B. Bergmann ;
P. Burian ; R. S. Brusa ; A. Camper ; F. Castelli ; G. Cerchiari ; R. Ciuryło ; G. Consolati ;
M. Doser ; K. Eliaszuk ; A. Giszczak; L. T. Glöggler ; Ł. Graczykowski ; M. Grosbart ; F. Guatieri ;
N. Gusakova ; F. Gustafsson ; S. Haider ; M. A. Janik; T. Januszek; G. Kasprowicz ; G. Khatri ;
Ł. Kłosowski ; G. Kornakov ; V. Krumins ; L. Lappo; A. Linek ; J. Malamant ; S. Mariazzi ;
L. Penasa ; V. Petracek; M. Piwiński ; S. Pospisil ; L. Povolo ; F. Prelz ; S. A. Rangwala ;
B. S. Rawat ; B. Rienäcker ; V. Rodin ; O. M. Røhne ; H. Sandaker ; P. Smolyanskiy ;
T. Sowiński ; D. Tefelski ; T. Vafeiadis ; C. P. Welsch ; T. Wolz ; M. Zawada ; N. Zurlo

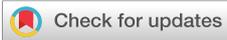

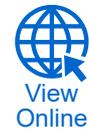
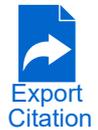

View Online    Export Citation

# TALOS (Total Automation of LabVIEW Operations for Science): A framework for autonomous control systems for complex experiments

M. Volponi,[1,2,3,a)] J. Zieliński,[4] T. Rauschendorfer,[1,5] S. Huck,[1,6] R. Caravita,[2,3] M. Auzins,[1,7] B. Bergmann,[8] P. Burian,[8] R. S. Brusa,[2,3] A. Camper,[9] F. Castelli,[10,11] G. Cerchiari,[12,13] R. Ciuryło,[14] G. Consolati,[10,15] M. Doser,[1] K. Eliaszuk,[4] A. Giszczak,[4] L. T. Glöggler,[1] Ł. Graczykowski,[4] M. Grosbart,[1] F. Guatieri,[2,3] N. Gusakova,[1,16] F. Gustafsson,[1] S. Haider,[1] M. A. Janik,[4] T. Januszek,[4] G. Kasprowicz,[17] G. Khatri,[1] Ł. Kłosowski,[14] G. Kornakov,[4] V. Krumins,[1,7] L. Lappo,[4] A. Linek,[14] J. Malamant,[1,9] S. Mariazzi,[2,3] L. Penasa,[2,3] V. Petracek,[18] M. Piwiński,[14] S. Pospisil,[8] L. Povolo,[2,3] F. Prelz,[10] S. A. Rangwala,[19] B. S. Rawat,[20,21] B. Rienäcker,[20] V. Rodin,[20] O. M. Røhne,[9] H. Sandaker,[9] P. Smolyanskiy,[8] T. Sowiński,[22] D. Tefelski,[4] T. Vafeiadis,[1] C. P. Welsch,[20,21] T. Wolz,[1] M. Zawada,[14] and N. Zurlo[23,24]

## AFFILIATIONS

[1] Physics Department, CERN, 1211 Geneva 23, Switzerland
[2] TIFPA/INFN Trento, via Sommarive 14, Povo, 38123 Trento, Italy
[3] Department of Physics, University of Trento, via Sommarive 14, Povo, 38123 Trento, Italy
[4] Warsaw University of Technology, Faculty of Physics, ul. Koszykowa 75, 00-662 Warsaw, Poland
[5] Felix Bloch Institute for Solid State Physics, Universität Leipzig, 04103 Leipzig, Germany
[6] Institute for Experimental Physics, Universität Hamburg, 22607 Hamburg, Germany
[7] University of Latvia, Department of Physics Raina boulevard 19, LV-1586 Riga, Latvia
[8] Institute of Experimental and Applied Physics, Czech Technical University in Prague, Husova 240/5, 11000 Prague 1, Czech Republic
[9] Department of Physics, University of Oslo, Sem Sælandsvei 24, 0371 Oslo, Norway
[10] INFN Milano, via Celoria 16, 20133 Milano, Italy
[11] Department of Physics "Aldo Pontremoli," University of Milano, via Celoria 16, 20133 Milano, Italy
[12] Institut für Experimentalphysik, University of Innsbruck, Technikerstrasse 25, 6020 Innsbruck, Austria
[13] Department of Physics, University of Siegen, Walter-Flex-Str. 3, 57068 Siegen, Germany
[14] Institute of Physics, Faculty of Physics, Astronomy, and Informatics, Nicolaus Copernicus University in Toruń, Grudziadzka 5, 87-100 Toruń, Poland
[15] Department of Aerospace Science and Technology, Politecnico di Milano, via La Masa 34, 20156 Milano, Italy
[16] Department of Physics, NTNU, Norwegian University of Science and Technology, Trondheim, Norway
[17] Warsaw University of Technology, Faculty of Electronics and Information Technology, ul. Nowowiejska 15/19, 00-665 Warsaw, Poland
[18] Czech Technical University, Prague, Brehová 7, 11519 Prague 1, Czech Republic
[19] Raman Research Institute, C. V. Raman Avenue, Sadashivanagar, Bangalore 560080, India
[20] Department of Physics, University of Liverpool, Liverpool L69 3BX, United Kingdom
[21] The Cockcroft Institute, Daresbury, Warrington WA4 4AD, United Kingdom
[22] Institute of Physics, Polish Academy of Sciences, Aleja Lotnikow 32/46, PL-02668 Warsaw, Poland
[23] INFN Pavia, via Bassi 6, 27100 Pavia, Italy

[24]Department of Civil, Environmental, Architectural Engineering and Mathematics, University of Brescia, via Branze 43, 25123 Brescia, Italy

[a]**Author to whom correspondence should be addressed:** marco.volponi@cern.ch

**ABSTRACT**

Modern physics experiments are frequently very complex, relying on multiple simultaneous events to happen in order to obtain the desired result. The experiment control system plays a central role in orchestrating the measurement setup: However, its development is often treated as secondary with respect to the hardware, its importance becoming evident only during the operational phase. Therefore, the AEḡIS (Antimatter Experiment: Gravity, Interferometry, Spectroscopy) collaboration has created a framework for easily coding control systems, specifically targeting atomic, quantum, and antimatter experiments. This framework, called Total Automation of LabVIEW Operations for Science (TALOS), unifies all the machines of the experiment in a single entity, thus enabling complex high-level decisions to be taken, and it is constituted by separate modules, called MicroServices, that run concurrently and asynchronously. This enhances the stability and reproducibility of the system while allowing for continuous integration and testing while the control system is running. The system demonstrated high stability and reproducibility, running completely unsupervised during the night and weekends of the data-taking campaigns. The results demonstrate the suitability of TALOS to manage an entire physics experiment in full autonomy: being open-source, experiments other than the AEḡIS experiment can benefit from it.

## I. INTRODUCTION

Physics experiments rate fairly high in the scales of size and complexity of information technology systems. To ensure the success of novel modern measurements, the synchronized coordinated behavior of multiple components is required. More often than not, the separate subsystems are constructed independently. While this enables parallelization during development, it frequently leads to difficulties during the integration and operation phases. In fact, each part is built according to the likes and experience of a single group of scientists: in some cases, e.g., for nuclear, atomic, and quantum physics experiments, the fields of expertise can be very different, ranging from lasers to electromagnetic traps, to ultra-cold and ultra-high vacuum, to various detection techniques. Mostly suffering from this is experiments' control systems' software. The term "control system software" refers to the combination of programs responsible for overseeing and managing the experimental apparatus.

Albeit interfaces for system integration might have been decided *a priori*, the various parts are coded with different paradigms, with different styles, and often even in different programming languages, leading to difficult subsequent iterations of unification. In addition, this limits the interplay among scientists and creates a great barrier to knowledge transfer: every time, a new way of reasoning has to be learned.

The AEḡIS experiment[1] is a perfect example of this situation. It is located at CERN's Antiproton Decelerator (AD) facility, and its main objective is to measure the gravitational deflection of a neutral antihydrogen beam. By nature, the AEḡIS experiment is a very complex and heterogeneous experiment.

The AEḡIS collaboration has already effectively generated antihydrogen atoms in pulsed mode.[2] During the establishment of antihydrogen formation, the constraints arising from the separate development of the different subsystems became evident: their integrability was not optimal, especially software-wise. With the increasing complexity of the experimental sequences necessary to make antihydrogen, the absence of programming constructs tailored to address this growth, coupled with a limited debugging ability of the running system and the restricted reusability of the previously scripted sequences, caused the operators' workload to rise enormously.

These limitations are not exclusive of the AEḡIS experiment: in fact, many solutions of control systems exist,[3–7] but none of them address simultaneously the problem of maintaining reliability, stability, and reproducibility of results while admitting the generality, scalability, and modularity necessary for the experiment to evolve and mutate, a crucial characteristic common to nuclear, atomic, and quantum physics experiments.[8]

Therefore, utilizing the knowledge gathered from previous experience, the AEḡIS collaboration decided to develop and integrate a new experiment control system, the CIRCUS (Computer Interface for Reliably Controlling, in an Unsupervised manner, Scientific experiments).[9] The foundation of this redesign lies in streamlining the complex experimental procedures through the standardization of established sub-procedures into libraries and in augmenting stability, reliability, and autonomy by an iterative process of implementation and debugging of the system.

Furthermore, recognizing the broad applicability of this approach, it was determined to develop a novel framework for control systems in atomic and quantum physics experiments and then construct the AEḡIS control system atop it. This framework, TALOS (Total Automation of LabVIEW Operations for Science), is the subject of this article.

This paper is structured as follows: in Sec. II, the TALOS framework is presented, its main structure and integration are described, and an explanation of how to use the framework to build a control system is given. The actual implementation for the AEgIS experiment and the results obtained so far with TALOS are presented in Secs. III and IV, respectively. Conclusions and future developments are outlined in Sec. V. In the Appendix, further details on the TALOS implementation are given.

## II. TALOS, THE FRAMEWORK

As stated in Sec. I, to address the common problems of nuclear, atomic, and quantum experiments' control systems, it has been decided to devise an experiment-agnostic framework that would serve as the common denominator to all the various components of the entire control system. It was required to be general enough so that every conceivably needed application could be written in it, inherently safe to operate under all circumstances, scalable, future-proof, and reliable. Moreover, to minimize the required time for operators to run it, it should be capable of great amount of automation.

From all these necessities, it is clear that the framework should be modular, easily expandable, self-checking, reliable, and able to operate concurrently on different machines harmoniously.

To meet the aforementioned necessities, the TALOS framework is founded upon two main pillars: "everything is a MicroService" (or, simply, μService) and the distributed system architecture.

The first concept is to divide the code into independent and autonomous parts, called μ*Services*, each with a clear scope and task. The μ*Services* are meant to be separate applications running asynchronously side-by-side, interacting with each other via a built-in messaging system. They all inherit from a common class, called *Father Of All* μServices (FOAM), which ensures, at the same time, code uniformity and maintainability: in fact, updates to the framework can be pushed "from the back," dealing at the same time with the back-end and the FOAM, but leaving the code, written by any user and specific to each μService, untouched.

The second concept comprises the idea of having an identical process, named *Guardian*, running on every machine, which monitors both the status of the other Guardians present on the network and the μServices running locally. Moreover, the Guardian supplies a common infrastructure to share messages and data between various μServices and among different computers. It unifies all the computers as a single distributed entity, and it enhances the stability, reliability, and safety of the system by having a distributed watchdog system so that no single computer becoming unresponsive can pass unnoticed. The unification of all the computers into a single entity is the key feature that enables the possibility to automate the entire system: in fact, the reaction to errors is global and independent of their origin, and high-level decisions depending on parameters generated by multiple computers are possible.

A schematic representation of the TALOS framework is shown in Fig. 1.

To fulfill these requirements, it has been decided to base the TALOS framework on the NI (*National Instruments Corp.*) LabVIEW[TM] Actor Framework, a LabVIEW implementation of the *Actor Model*, of which both are briefly introduced in Subsection II A.

### A. The Actor Model and NI LabVIEW Actor Framework

The Actor Model[10] is a versatile computational model for designing concurrent and distributed systems. It facilitates the organization and structuring of software components to enhance scalability, fault tolerance, and responsiveness. It uses autonomous entities called actors, each with its own state, behavior, and message inbox. Actors process messages sequentially from their queue, typically in a First-In-First-Out (FIFO) order, and communicate by passing messages to each other. They can create new actors dynamically, enabling hierarchical systems. Actors work asynchronously, independently processing messages without blocking one another. This feature is especially valuable for highly concurrent and responsive systems, efficiently utilizing resources and handling substantial workloads. Moreover, it enhances fault tolerance, as errors or failures in one actor do not disrupt the entire system, and the isolation of actors simplifies testing and development.

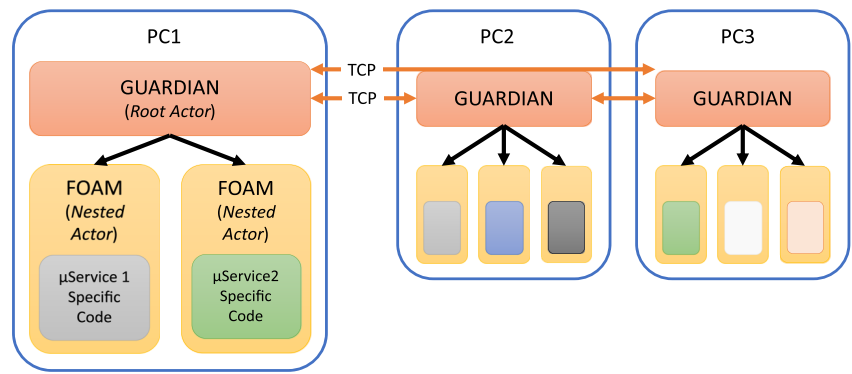

**FIG. 1.** The TALOS framework structure. The framework is built on the *Actor Model*, where on each machine, the Guardian acts as a unique *Root Actor* and all the μServices are *Nested Actors* of it. On each PC, an instance of Guardian is launched, which in turn launches all the μServices that should run on that machine. The Guardians maintain the communication and watchdog layers that unify the distributed system into a single entity (enabling, for example, each μService to message any other μService, regardless of the computers they are running on, possibly different), and they monitor the local μServices for unresponsiveness. The interaction between each Guardian and its own μServices is coded into FOAM, the parent class of every μService.

In the LabVIEW programming language,[11] a native implementation of the Actor Model is offered, the *Actor Framework*. Actors in LabVIEW are represented by individual Virtual Instruments (VIs) that encapsulate both data and behavior. The Actor Framework provides a structured way to create, manage, and coordinate these actors, making it easier to develop complex systems. The framework manages the execution of actors and ensures that messages are processed in a controlled and synchronized manner; moreover, it facilitates the handling of exceptions, by enabling each actor's reaction to them and methods for error transmission between actors. In addition, the Actor Framework promotes modularity and code reusability, by leveraging the power of object-oriented programming (each actor is a class).

The characteristics of the Actor Model (and the Actor Framework) made it the ideal starting point for the development of the control system, following the two pillars outlined at the beginning of Sec. II. In fact, it is by nature, suitable for building distributed systems, and the actor features (modularity, independence, and asynchronicity) are exactly what was envisaged for the μServices.

### B. TALOS structure

As mentioned before, the TALOS framework is based on the LabVIEW Actor Framework, since it is designed for the implementation of multiple asynchronously interacting processes. The actors' hierarchy is simple: on each PC, the Guardian acts as a *Root Actor*, and all the μServices are *Nested Actors* of it (see Fig. 1). This flat organization simplifies the management of the various components, making the system more resistant to failures, since stopping one μService does not affect its siblings.

The power of class inheritance is used for the μServices since FOAM is the parent actor of all the μServices in the system. It both masks the Actor Framework complexity to the end-developers and manages the interaction of the μService with the Guardian. Moreover, it guarantees the maintainability of the system: in fact, all the μService-specific codes inserted into each child of FOAM remain untouched during updates to TALOS, since only the FOAM specific VIs are modified.

The communication between Guardians and μServices relies on the *Transmission Control Protocol* (TCP).[12] It consists of using two separate pairs of μServices, the *TCP Listener* and *TCP Writer*, to instantiate two tunnels between each two Guardians where data flow in a single direction, allowing parallel and asynchronous message handling. This solution is also used for the communication between TALOS and the FPGAs, as explained in Sec. II C.

Another salient characteristic of the μService structure is that the same response can be triggered by a button on its GUI or by an external message: this guarantees that, after testing the μService manually from its GUI, the expected behavior will be maintained during its programmatic use.

In the Appendix, more details on the implementation of TALOS can be found, ranging from an explanation of the Guardian watchdogs system (A 1) to the μService VI scheme (A 2).

### C. Sinara and ARTIQ

The timing precision needed in modern physics experiments (typically, at least, on the order of the nanoseconds) cannot be handled by PC Operating Systems (OS) without real-time hardware and OS extensions whose complexity often makes it preferable to delegate time-critical operation to one or more *ad hoc* Field Programmable Gate Arrays (FPGAs) or programmable Systems On-Chip (SOC). In the AEg̅IS experiment, the Sinara[13] ecosystem has been chosen as the base of the (new) control system electronics, which comes with modules for both FPGA and SOC approach.[14]

Sinara comprises a varied collection of open-source hardware components initially designed for quantum information experiments. The Sinara hardware is suited for handling complex, time-sensitive, and spatially limited experiments since it delivers small, modular, reproducible, and dependable electronics. In addition, its characteristics future-proof the maintainability of the system.

This electronics is organized in rack-mounted crates, where a single controller, called Kasli, directs multiple (up to 12) modules with different characteristics [for example, digital input/output (I/O) units, fast digital-to-analogue converter (DAC) modules, and 1 MHz high-voltage amplifiers]. A fiber connection between controllers enables them to be joined and used in a master-satellite fashion.

This FPGA is programmed with a custom programming language called ARTIQ (Advanced Real-Time Infrastructure for Quantum physics).[15] ARTIQ is a high-level programming language based on Python, so routines for experiments running on Kasli-controlled electronics are simply scripts, which call dedicated functions to interact with the hardware. This language simplifies enormously[16] the learning curve for adopting it, and it renders possible the utilization of custom-defined libraries, which enables both to make the code extremely self-explanatory and to foster code reuse by minimizing duplication.

Typically, Kasli is operated through a computer shell, where a script containing a sequence of instructions is manually sent to Kasli. While this approach facilitates debugging, it requires an operator to monitor the operations and control script submission.

Therefore, a specific μService has been created to manage the execution of the scripts of Kasli, the *Kasli Wrapper*. This μService both encapsulates the ARTIQ shell to internalize the handling of the schedule of scripts inside TALOS, giving it full control over its flow (see Sec. II E) and intercepts the terminal-based commands, to be able to catch also the low-level exceptions raised by Kasli, facilitating all-rounded error management. Moreover, upon each script completion, it forwards to the *Monkey* (which is the μService managing the high-level control: see Sec. II E), the **B**asic qu**A**lity **N**otification **A**fter the e**N**d of an **A**ction (or BANANA), summarizing the script execution result, to ensure proper decision taking.

The communication protocol outlined in Sec. II B further strengthens the integration. In fact, the Kasli normal user interface is a command-line terminal, where a user controls the outputs and provides an eventual input. In the AEg̅IS experiment, though, Kasli is the actual orchestrator of the operations at the nanosecond level, so it necessitates a bi-directional exchange with computers, to send commands and receive inputs. Re-routing the terminal communications proved unfeasible as it can interfere with the real-time operation of the FPGA. The double, asynchronous TCP tunnel between TALOS and the Kasli allows the Kasli to communicate without losing the nanosecond-precise schedule of the sequence of operations.

To ease the exchange between Kasli and the various μServices, a custom library for ARTIQ, called *TCP Library*, has been created. It hides the complexity of the TCP interface inside a few functions to be called from the importing script. Here follows an example of the syntax of a function to send a message, with data, to a specific μService.

```
def NewFunctionName(parameters):
    TCP_Send("<MicroService-Name>;" +
        "<Message-Case-Name>;" +
        "<Data1>;<Data2>;<etc.>")
```

It has to be noted that Sinara Kasli is the FPGA integrated by default into TALOS. However, by overriding the *Kasli Wrapper*, any other FPGA capable of TCP communication can be accommodated similarly.

### D. ALPACA

In TALOS, an interface for data analysis engines is present. They have to provide two functionalities: returning the value of a pre-defined observable, given the experimental conditions, and suggesting points in a pre-defined parameter space, based on the feedback of live data taken. This interface extends the capabilities of TALOS beyond scheduling experiments, giving the possibility to perform, in full autonomy, optimizations of observables over a given parameter space, which renders the control system capable also of taking decisions driven by online data feedback, as shown in Secs. II F 3 and II F 4.

In the AE͞gIS experiment, TALOS has been interfaced with *All Python Analysis Code of AE͞gIS* (ALPACA), the data analysis framework for the AE͞gIS experiment written in Python.[9] Its linear and scalable architecture allows for easy integration of new observables depending on the envisioned experiments, and the automated deployment makes these observables accessible to the control system without any user interaction. Most importantly, ALPACA implements a Bayesian optimizer, capable of suggesting points to explore in set parameter space to quickly converge one (or more) observable toward an optimum value.

ALPACA's optimizer uses *Scikit-optimize*, utilizing Gaussian processes as a surrogate model, and the GP-Hedge algorithm for choosing the best of the three implemented acquisition functions: *Expected Improvement*, *Probability of Improvement*, and *Lower Confidence Bound*.[17]

### E. Notable μServices

μServices are coded in the system as needed by the experiment. Nevertheless, some special μServices are already integrated into TALOS because they perform functions generally useful to every experiment: examples are the aforementioned μServices responsible for communication. The most notable μServices integrated into TALOS are as follows:

- **Error Manager:** A key part of the automation provided by TALOS lies in the distributed error management system. The system responds consistently to errors arising from every machine in the experiment: this is obtained by contextualizing each error, substituting it with one from a pre-defined list, with a precise *criticality code* associated. Each criticality code corresponds to an action for the *Monkey* (see below).
- **The Scheduler:** This μService provides the user with a Graphical User Interface (GUI) for defining schedules of experiments, a feature fundamental for an autonomous control system. Each schedule is defined as a series of Schedule Blocks (or *SBlocks*), each consisting of an experimental script and a series of parameters. Two types of SBlocks can be defined: *Scan SBlocks*, to perform the same experiment multiple times, spanning over a (multidimensional) parameter space; and *Optimization SBlocks*, which autonomously search for the best point in a parameter space to optimize the value of a pre-defined observable, leveraging the interface with ALPACA.
- **The Monkey:** It is the core of the automation of TALOS. It executes the schedule of experiments defined with the Scheduler (see above), and, upon each script completion, it decides the action to be taken based on the outcome of the finished script (i.e., the BANANA) and the eventual errors that occurred during its execution. The five possible actions are *Continue*, *Retry*, *Skip*, *Stop*, and *Abort*. It also manages the communication with ALPACA for the optimization workflow (as shown in Fig. 4).
- **The Tamer:** TALOS enables the use of multiple Kasli controllers in parallel, through the simultaneous operation of several instances of *Monkeys* and *Kasli Wrappers*, a pair for each Kasli controller used. To ensure and control the proper flow of data, the *Tamer* μService is used as a distributor: it receives the BANANA messages from the *Kasli Wrappers* and re-routes them to the addressed *Monkeys*. This scheme allows for synchronous and asynchronous execution of scripts in parallel mode (see Sec. II F 2).
- **Detector Manager:** Modern physics experiment utilizes multiple detectors to meet their scientific goals. Albeit very different, their operations can be schematized in: configuration, data acquisition, and data saving. With this standardization, a generic μService—called *Detector Manager*—was coded, to quicken the integration of detectors in the experiment control system by having pre-coded most of the μService logic, necessitating only the specification of the code to communicate with the device.
- **DAQ Manager:** This μService manages the interface with the Data AcQuisition system (DAQ), a fundamental part of every scientific experiment (see Fig. 4). Every DAQ with commands for starting it, stopping it, and sending data can be integrated into TALOS: afterward, every μService in the system can send data to the DAQ simply using a dedicated VI.
- **Kasli Wrapper:** It manages the interaction with the FPGA (see Sec. II C).

Further details on these μServices can be found in the Appendix.

### F. Autonomous operation

The most notable characteristic of TALOS is its ability to handle the execution of entire schedules of experiments without the need for human supervision. This was one of the main goals from the start, so the entire system was specifically designed for this task. This capability is crucial to maximizing the amount of data taken while minimizing the operators' time devoted to caring for the machine; moreover, it increases the repeatability of experiments by minimizing random events and human errors.

#### 1. A boat with two captains

To safely and reliably execute a schedule of antimatter physics experiments, both real-time system status awareness and nanosecond precision timing are essential. These two properties cannot be satisfied by either TALOS or Kasli alone: in fact, TALOS has the system overview, but it is limited to the millisecond-level scheduling precision of regular, non-real-time PCs; in contrast, Kasli offers nanosecond-precision, but its scope is limited to its internal status and the digital and analog input line voltages. Therefore, the control system resembles a boat with two captains, periodically switching the helm control depending on the needs.

Initially, when the schedule of runs begins, TALOS assumes control. It verifies the correct functioning of each μService, and then, it sends the first script to Kasli. Here, the helm control shifts to Kasli, which executes the script, while TALOS acts as a "slave," redirecting messages to μServices (TALOS only intervenes in the event of an `ABORT`, halting the execution and entering *Safe Mode*). At script completion, Kasli returns control to TALOS, which assesses the return code and any eventual error to determine the next action, which typically involves submitting another script to Kasli. The process repeats until the schedule is completed or the execution halts due to an error or user intervention.

#### 2. Automation flow

The system allows for three modes of operation (depending on how and how many Kaslis are controlled): sequential (standard), asynchronous (parallel), and synchronous (parallel). The automation flow begins with the schedule being defined by the user in the *Scheduler* and sent to the *Tamer*, which starts deciding how many *Monkeys* and *Kasli Wrappers*[18] need to be running. Subsequently, after having verified that all Guardians and μServices are ready, the *Tamer* propagates to each *Monkey* the corresponding part of the schedule, and the first Run starts. This procedure is a common start that is independent of the operation mode. Each *Monkey* starts executing its schedule and waits for the Run's outcome.

In the sequential operation mode, only one *Monkey* is running; nevertheless, the (single) schedule can have scripts designated for different Kaslis. The *Monkey* runs the scripts one by one, waiting for a `BANANA` message (i.e., script termination) before starting a new script, even if they are designated for different Kaslis. A scheme of the automation flow for the sequential operation mode is visible in Fig. 2.

The asynchronous parallel operation mode is analogous to the sequential mode but with multiple *Monkeys* running multiple schedules simultaneously [see Fig. 3(a)]. Each schedule is a set of scripts for a specific Kasli, assigned to a dedicated *Monkey*, to have a 1–1 correspondence. Each *Monkey* executes its own schedule independently from the other *Monkeys*, allowing multiple scripts to run in parallel, asynchronously, on different Kaslis.

In the synchronous parallel operation mode, the main goal is to start all the scripts belonging to different Kaslis at the same time: therefore, each *Monkey* needs to know the status and outcome of all its siblings. This functionality is achieved by using the *Tamer* as an execution barrier and a response collector: it gathers all the outcomes of the various *Monkeys*' script checks and the `BANANA` messages, and redistributes the summary to every *Monkey* simultaneously, to ensure their synchronicity. A scheme is shown in Fig. 3(b).

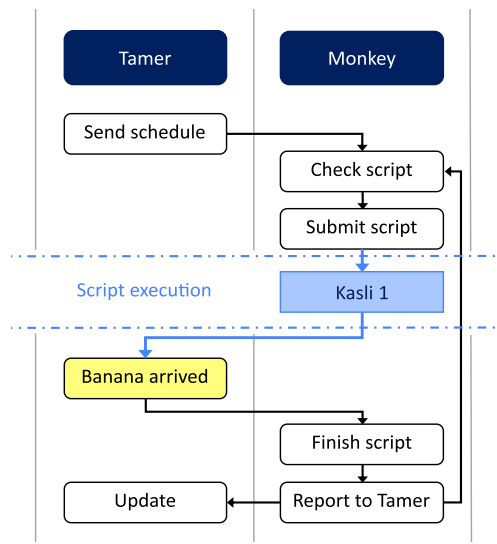

**FIG. 2.** Scheme of the autonomous operation flow in the sequential operation mode (i.e., single *Monkey*). For simplicity, only one *Kasli* has been depicted, even if multiple ones can be used, but not in parallel.

The user defines the mode of operation in the *Scheduler* thanks to a "synchronization mask," which allows the system to synchronize multiple Kaslis while running others asynchronously. The results of the evaluation of synchronicity are shown in Sec. IV E.

#### 3. Automatic parameter optimization

Although the level of user independence depicted until this point is already high, the automation is limited to reacting to errors and external events. This already had a huge positive impact on running the experiments at AE͞gIS, but it was decided to go further. ALPACA, the AE͞gIS analysis framework, was interfaced with TALOS, to empower the latter with the ability to change the parameters of the scripts based on the results of the experiments performed previously.

With the *Scheduler*, an *Optimization SBlock* can be defined (see Sec. II E), consisting of an experimental script, a parameter space to explore (with a starting point), and an observable to optimize. This script is executed normally (using the starting parameters provided by the user), and, upon completion, ALPACA suggests the next point to explore in the parameter space, based on the data just acquired. The *Monkey* then re-executes the script, using the ALPACA feedback. This procedure is iterated until ALPACA declares that the optimization has converged, or the maximum number of iterations allowed by the user is reached. The rest of the schedule is then executed normally.

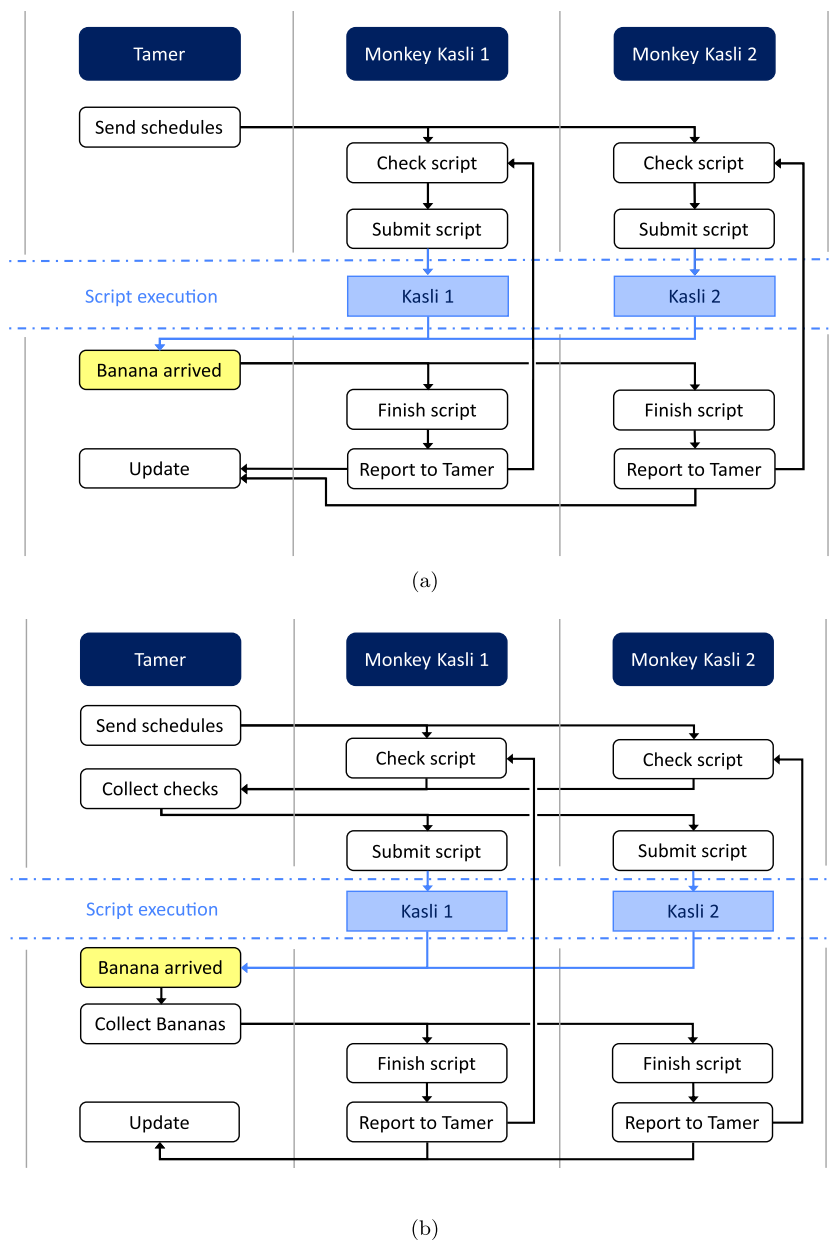

**FIG. 3.** Schemes of the autonomous operation flow in the parallel asynchronous mode (a) or parallel synchronous mode (b). In both cases, an arbitrary number of Kaslis can be coordinated.

This new operational approach fundamentally changes how experiments are conducted. Instead of manual exploration of a wide parameter space followed by data analysis, the optimization problem is now embedded in the script, allowing TALOS to autonomously find the optimum. This approach offers faster convergence than the traditional grid scan, especially in multi-parameter optimization, where ALPACA's Bayesian optimizer scales linearly with the parameter space dimensions, as opposed to the exponential scaling of grid scans (see Sec. IV B). This removes the necessity of assuming, in multi-dimensional problems, the orthogonality of parameters, potentially leading to better operational settings.

Complex tasks, like the laser calibration needed at AEḡIS, are completed more swiftly by TALOS, taking less than one hour, compared to the previous manual process that required several hours from an expert. Moreover, once calibration procedures are automated, they can be run periodically, to maximize system performance over time.

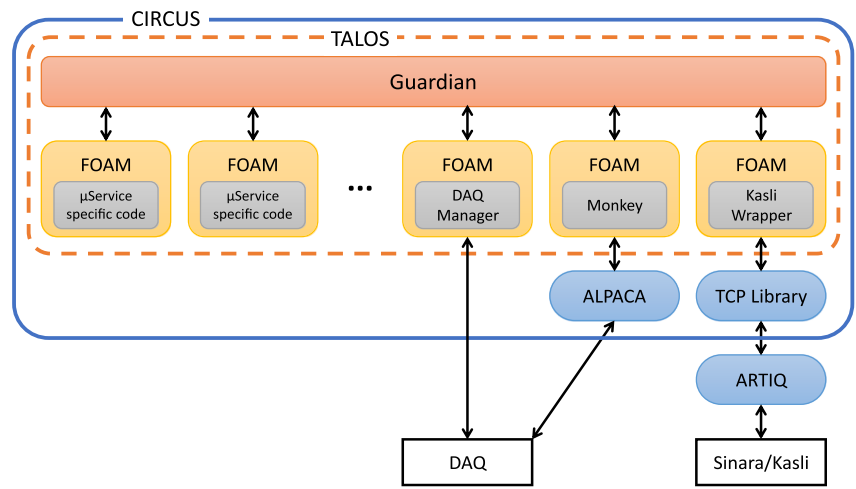

**FIG. 4.** Scheme of CIRCUS and TALOS, depicting the relation between Guardian, μServices, ARTIQ/Sinara, ALPACA, and the DAQ. The actor-based structure of TALOS is the base of the interface between the Guardian and the μServices, encapsulated into the FOAM: in each μService, only the code for the specific functionalities needs to be implemented. Three notable μServices are explicitly mentioned: the *DAQ Manager*, controlling the interaction with the data AcQuisition system; the *Monkey*, the core of the automation and managing the interface with ALPACA; and the *Kasli Wrapper*, carrying the interaction with Sinara/Kasli through the *TCP Library* and ARTIQ.

### 4. Quality of Run assessment

Another use of the TALOS-ALPACA integration is to assess the quality of a run. In the preparation of a schedule, checks can be defined (in the form of *Observable* < / ≤ / = / ≥ / > *value*) to determine the data quality upon script completion. If the overall return code is *Continue*, the *Monkey* contacts ALPACA to retrieve the observable values for testing. If any checks fail, the Run is marked as unsatisfactory and retried.

This addition allows the system to pre-filter data, preventing the need for manual exclusion during the analysis stage and subsequent re-taking of points. This is particularly advantageous when combined with the optimizer, as pre-filtering avoids biasing the system during its autonomous search for optimal parameter values.

### G. CIRCUS: How to use TALOS

TALOS is intended to be the underlying framework upon which the control system of an experiment can be based.[19] It is thought to be used as the core engine of the CIRCUS control system, where TALOS comes pre-compiled as a LabVIEW binary *Packed Project Library*, pre-loaded into a LabVIEW example project, also containing a template for μServices. The project configuration file, the custom error file, and *Startup.bat* come bundled with it.

Using the template given, new μServices can be easily added to the CIRCUS, and the configuration and custom error files can be modified as needed. In this manner, CIRCUS can be modeled to meet the necessities of the experiment to control, leveraging the capabilities of TALOS outlined in this article.

CIRCUS also comprises ALPACA and the libraries facilitating the interface between the FPGA (by default, Kasli) and TALOS. A scheme of the relations between CIRCUS and TALOS, and the external parts of the experiment, is shown in Fig. 4.

The CIRCUS control system is available open-source in a *Git* repository (DOI: 10.5281/zenodo.10371799).

A screenshot of the CIRCUS control system executing a double schedule of scripts is given in Appendix 6.

## III. REAL USE CASE: AEḡIS CONTROL SYSTEM

### A. The AEḡIS experiment

The AEḡIS experiment is located at the Antimatter Factory, hosted at CERN (European Organization for Nuclear Research). The main goal of the experiment is to measure directly, with high precision, the fall of antimatter in the earth's gravitational field. Specifically, the aim is to measure the vertical displacement—caused by gravity or gravity-like interactions—of a pulsed beam of antihydrogen. The scientific interest of such a measurement is to find a possible deviation from the symmetry between matter and antimatter, which could explain its scarcity in the known universe. Although limits from astrophysical sources and time-scaling effects[20] and indirect measurements have already been performed,[21] a direct measurement is of great interest because of its model-independency. Currently, the best direct measurement is the one from ALPHA:[22] while a huge milestone in the research sector, its precision is more than one order of magnitude away from where a deviation is reasonably expected (i.e., below 1%).

The AEḡIS experimental apparatus consists of two collinear Penning–Malmberg traps hosted inside the same cryostat, a positron line able to bring positrons to the positron-to-positronium ($e^+ \to Ps$) converter, and a system of multiple lasers to excite Ps in order to form antihydrogen. The first trap has a magnetic field of 5 T and electrodes that can go up to 15 kV. It is used to capture the antiprotons ($\bar{p}$) coming from ELENA (Extra Low ENergy Antiproton ring—it is the new decelerator in operation in the AD, feeding antiprotons to the experiments at 100 keV). These are then sympathetically cooled with electrons and transferred to the second trap, called the production trap, where they are further cooled and stored to make antihydrogen. This second trap is characterized by having a magnetic field of 1 T and electrodes limited to 200 V. The positron line consists of a radioactive source of positrons ($^{22}$Na), a moderator, an accumulator to bunch them, and a transfer line for bringing the positron bunch in the production trap, close to the $\bar{p}$ plasma, with an energy of 5 keV. Here, the positrons impact onto the $e^+ \rightarrow$ Ps nano-channeled silica converter, to invest the $\bar{p}$ plasma with Ps atoms thus formed. To enable a charge-exchange reaction between the two, the positronium atoms, while flying, are excited from the ground state, in which they are produced, to a high ($n$ = 17–25) Rydberg state (Ps*) thanks to a series of two laser pulses (as shown in Fig. 5). The antihydrogen produced is detected via its decay products using scintillator slabs covering the entire length of the apparatus.

During AE$\bar{g}$IS Phase 1, lasted until 2018, antihydrogen production was demonstrated,[2] at the price of a great effort from the scientists. In fact, the experiment was composed of a multitude of independent subsystems coordinated by an overgrown common control system. While this has been precious during the development, enabling work parallelization, it became a limitation when a synchronized orchestration of the entire system became needed for antihydrogen production.

In 2019, with CERN's Long Shutdown 2 (LS2), AE$\bar{g}$IS entered its Phase 2, with the aim of enhancing antihydrogen production by 2 to 3 orders of magnitudes, to produce it 10 times colder, and to test the first antimatter gravimeter.[23] To achieve these results, a series of major upgrades have been performed on the experimental apparatus.[24,25] Among these, there have been changes to how the Ps illuminates the $\bar{p}$ plasma (from orthogonal to collinear, removing the limit on the Ps Rydberg level imposed by the field ionization due to the motional Stark effect), a completely redesigned formation trap, efficiency and stability improvements to the positronium line, a higher-yield $e^+ \rightarrow$ Ps converter, a more powerful laser system, and the migration of most of the core electronics and software control system to ARTIQ/Sinara and TALOS.

The upgrades to the apparatus were also warranted by the operating mode of the new ELENA[26] (Extra Low ENergy Antiproton) ring: antiprotons were formerly provided in shifts of 8 h per experiment, with an energy of 5.3 MeV, while now ELENA supplies $\bar{p}$ at an energy of 100 keV, every 2 min, 24 h/day. The lower energy of the antiprotons has significantly improved the AE$\bar{g}$IS' $\bar{p}$ trapping efficiency. However, with the previous control system, which required extensive supervision, the new $\bar{p}$ delivery rate would not have been sustainable, since it would have placed a heavy burden on operators, potentially limiting the utilization of the extended beam time. Transitioning to a highly automated control system was essential to maximize data collection and reduce user oversight, to allocate more time to physics and development.

### B. CIRCUS and TALOS at AE$\bar{g}$IS

In the AE$\bar{g}$IS experiment, multiple different subsystems are present: lasers, vacuum systems, particle traps with high voltage electrodes, scintillators, detectors, actuators, and many more. At the same time, the current operation rates imposed by the introduction of ELENA mean that the system needs to be operational 24 h a day. Utilizing the CIRCUS, based on TALOS, all the subsystems can be unified under one entity that autonomously controls and overviews the experiment's operation, maximizing the beam time taken. Furthermore, it provides the synchronization and the orchestration of operations between all submodules essential for high efficiency $\bar{p}$ trapping and antihydrogen production, enhancing its formation and consequently improving and enabling new physics results.

In 2021 and 2022, the AE$\bar{g}$IS control system has been progressively migrated to use the CIRCUS. The partitioning of the system into μServices allowed for incremental development, where each portion of the old code was ported into a new μService, rigorously tested, and debugged before integration into the live control system. This approach minimized the downtime introduced by code issues and facilitated rapid debugging. In fact, if a problem emerged with a newly added μService, the control system could be swiftly reverted to the previous working state by removing it, and the bug was known to be in the new μService, simplifying the debugging process.

At the time of writing, the new control system encompasses a network of six computers, running a collection of 120 μServices, some of which are different instances of the 42 unique μServices coded in the project. These manage several pieces of hardware, including three cameras, three different spectrometer types, two laser crystal heaters, seven actuators for laser-optic components, two oscilloscopes, the electron gun, the high voltage generator, the pulser, the rotating wall generator, and other devices. The stability of the system has led to the integration, at the end of 2022, of the environmental control system as a μService: it is one of the most critical pieces of software of the AE$\bar{g}$IS experiment, since it checks and maintains the status of the vacuum and the cryogenic temperature of the entire experiment, upon which depends the life of AE$\bar{g}$IS superconducting magnets. In fact, a warming of the magnets during operation can easily cause a quench, potentially fatal to the magnets. More than 500 custom errors have been defined, and the system has been online since August 2021.

The AE$\bar{g}$IS experiment has not only benefited from the new control system in terms of purely enhanced performances but especially from new capabilities that were unthinkable before. A good example of this is the *ELENA Interface*, our interface with the accelerator, which enables us both to control the status of the beam—and retry a measurement in case of antiproton unavailability—and to control some of its parameters. This feature, in turn, led to the automation of tasks that were previously only possible manually, like the beam steering (more on this in Sec. IV).

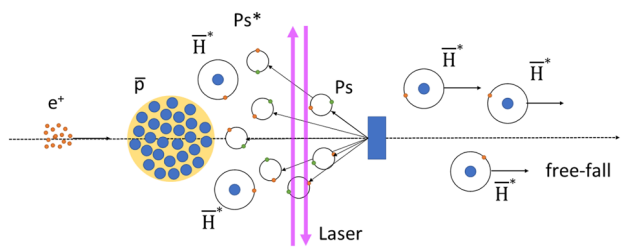

**FIG. 5.** Scheme of the antihydrogen production technique used at AE$\bar{g}$IS. It leverages the charge-exchange reaction, where a trapped cold plasma of antiprotons is invested by a cloud of Rydberg-excited positronium atoms, created by positrons impinging on a nano-channeled silica converter and subsequently excited by a series of two lasers.

Another example is the *Telegram Bot*, which enables the operators in remote to quickly see on their phones the status of the running schedule and even to subscribe to receive periodic updates (e.g., every 30 min).

However, by far, the biggest addition in terms of capabilities is the "smart" automation mentioned in Sec. II F and realized through the close relation between all the μServices described in Sec. II E. It reduces enormously the pressure on the operators, while also enhancing the repeatability of the experiments by lowering the possibility for human errors. This feature was fundamental in achieving all the results presented in Sec. IV.

CIRCUS is the subject of a dedicated publication.[9]

## IV. RESULTS

Thanks to TALOS, the insertion of new features and interfaces with additional hardware devices was demonstrated to be possible in parallel to the data acquisition, minimizing system downtime due to debugging during physics campaigns. This mode of operation has proven crucial because the system was always operable, even during its development.

### A. Physics results

The flexibility and modularity allowed the system to be operative in a very short time. During the first $\bar{p}$ campaign, in 2021, we managed to successfully trap antiprotons in the experiment with TALOS in less than a week, several times faster than with the previous system. The possibility of running automated scans (overnight, while debugging and development took place in the daytime) was exploited immediately to explore the effect of the trap closure time on the trapping efficiency and to characterize the energy of the antiprotons thus captured. The results of this scan are visible in Fig. 6.

The antiproton campaign of 2022 demonstrated the reproducibility that TALOS enables. For example, the capture of the antiprotons was achieved on the first day of the beam taken, simply running the PbarCatchNDump.py procedure developed the year before. The script controls the potentials on the trapping electrodes, which need to be synchronized with the triggers received from the ELENA, signaling the arrival of antiprotons to the experiment. Next, the script needs to prepare all the detectors (MCP and scintillators) for releasing antiprotons from the trap and measuring the annihilation rates to establish the trapping efficiency of the experiment. The procedure receives and sends triggers using a Kasli controller, while the electrodes' power supply, the MCP detector, and other environment-dependent hardware are controlled via μServices. Furthermore, the implementation of the *ELENA Interface* (see Sec. III B) enabled both further increases in the trapping efficiency, by automatically scanning over the beam parameters to find the best ones (see Fig. 7), and a higher stability and uptime, by making TALOS react to external events like *no beam*, *valve closed*, *empty shot*, and *beam stopper in* (see Sec. IV C).

All these improvements have contributed toward achieving the record of $\bar{p}$ trapping efficiency.[24] Preliminary analyses point forward a systematic trapping of ~$3.4 \cdot 10^6$ $\bar{p}$ per bunch, with efficiency around ~70 %.

### B. Autonomous parameter optimization

As explained in Sec. II F 3, the integration of ALPACA in TALOS has rendered the control system able to autonomously search the best parameter setpoint to optimize a given observable. A good example is the optimization of the beam steering to maximize the number of antiprotons trapped, a task that has to be performed repeatedly during every antiproton beam time. The parameter space is made up of the real numbers of the vertical and horizontal offsets and angles of the incoming antiproton bunch relative to the axis of the injection line, which means optimization of a four-dimensional space. The number of trapped antiprotons is proportional to the number of annihilation events detected by the installed scintillator detectors upon a "dump" of the content of the trap toward one end.

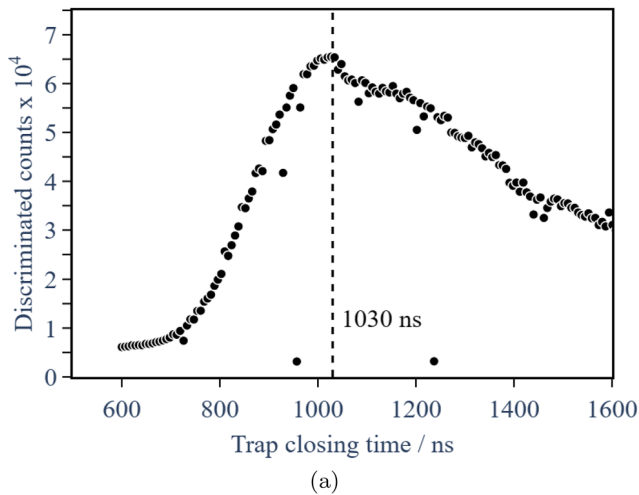 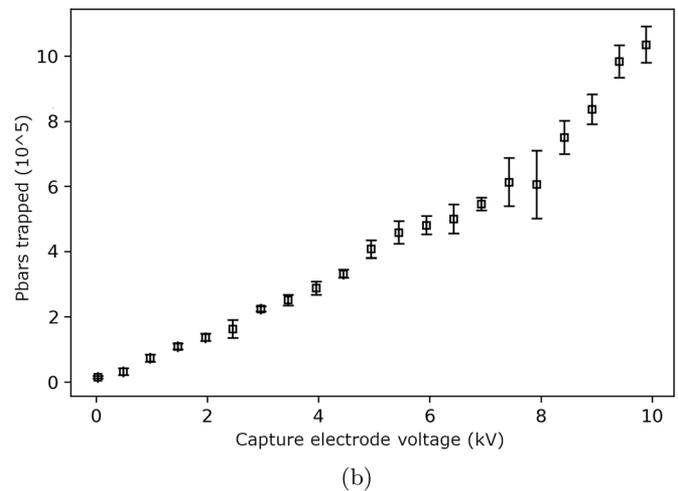

**FIG. 6.** (a) Graph showing the number of antiprotons captured vs the closure timing of the trap. It clearly shows the presence of a best working point. Closing too fast lets some antiprotons out, and, conversely, closing too slow lets some antiprotons escape after the bounce on the second electrode. (b) Graph showing the number of antiprotons captured varying the potential of the catching electrodes. This scan characterizes the energy profile of the $\bar{p}$ ′ s passing through the degrader, and their ratio is in good accordance with our GEANT4 simulations.

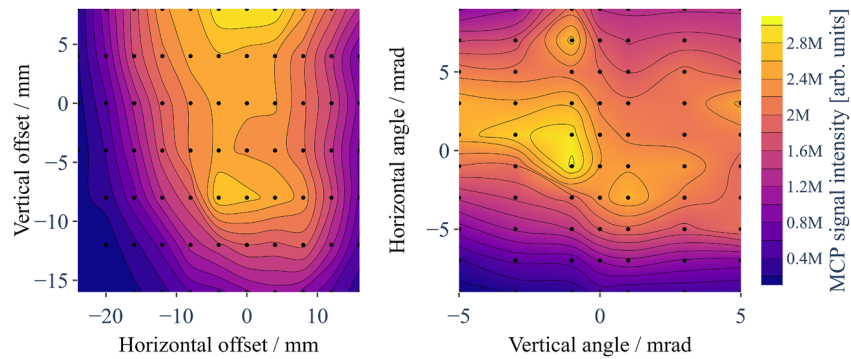

**FIG. 7.** Two graphs show the results of the scan over the horizontal and vertical displacements of the antiproton beam (on the left) and the horizontal and vertical angles (on the right). The color represents the intensity of the signal obtained on the MCP from the annihilations of the trapped antiprotons. The parameter space has been organized in this way, assuming that displacements and angles have independent effects, not for physics reasons, but because scanning over the full parameter space would have been impossible time-wise (10 steps per dimension 4 dimensions × 5 min of duration of the script ≈35 days!).

Figure 8(a) shows the largest number of observed annihilation events over the course of 101 consecutive measurements with the parameters suggested by the Bayesian optimizer. Highlighted in yellow are the initial 30 runs used to randomly explore the parameter space. For benchmark purposes, we defined a reference convergence criterion, which is evaluated as soon as the number of conducted experiments exceeds the initial exploration experiments,

$$\left| \frac{\sigma_{best10}}{\mu_{best10}} \right| < \delta. \qquad (1)$$

Here, $\mu$ and $\sigma$ reflect the mean and standard deviation of the top ten measurements with the highest fitness function score from all runs of the same experiment conducted for this optimization session. The fitness function depends on the selected optimization strategy and takes optimization parameters as the arguments. $\delta$ is to be chosen by the experimenters.

The convergence criterion was pre-defined with $\delta = 0.05$ and was met after 63 experiments. Given the relative improvement of additional ~23% throughout the next 38 experiments, we find that the convergence criterion (1) could have been more strict.

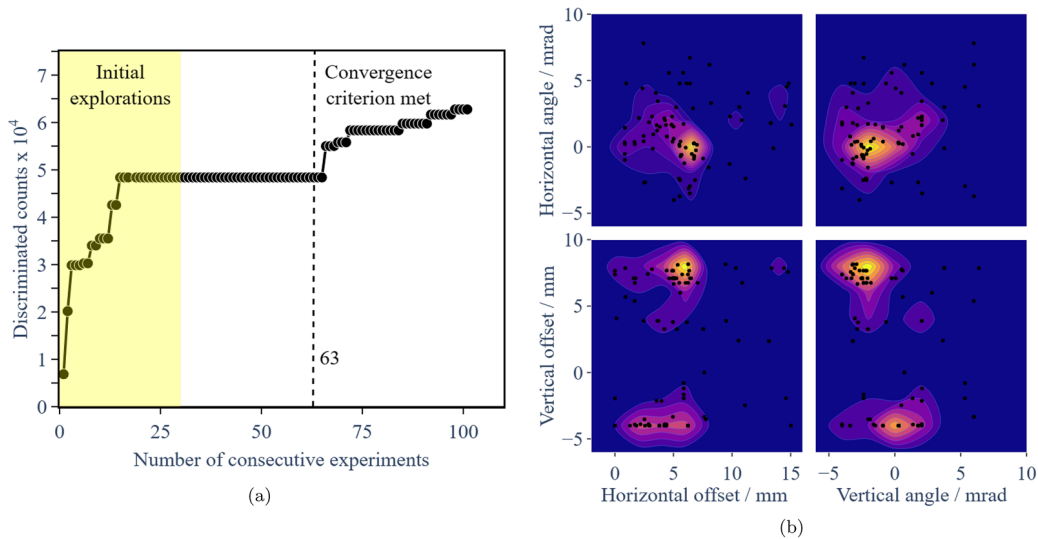

**FIG. 8.** Results of the beam steering optimization: (a) the convergence plot of the highest observed number of annihilation events and (b) the density plots of the evaluations in the four-parameter space dimensions.

The highest value throughout the 101 experiments was observed for the following parameters: horizontal offset: 6.23 mm, vertical offset: 7.38 mm, horizontal angle: −0.90 mrad, and vertical angle: −3.17 mrad, which matches with the density of evaluations in the parameter space shown in Fig. 8(b). This result is in excellent agreement with the results from the previous year. The difference in the shape of the graphs between Figs. 7 and 8(b) comes from the difference in the methods used for obtaining the data. The first plots were obtained using a scan where the full parameter space is mapped and plotted. In the second approach, the optimizer selects points based on previous results, closing on the optimum parameterization and mapping only areas near global and local optima in detail. The number of experiments required to reach a result comparable to the manual scan performed in 2022 shows a speed increase of about 146%. Actually, the performance increase is much higher, since the Bayesian optimizer was operated without restrictions on the parameter space, eliminating the previous need to conduct scans in multiple sub-spaces.

The evolution of the best-observed setting shown in Fig. 9(a) shows that the initial random exploration of the one-dimensional parameter space already got almost the best setting. The convergence criterion (1), using a more rigorous threshold $\delta = 0.02$, was met after 31 experiments, which is in agreement with the scan performed in 2022 with a total of 140 experiments. This corresponds to a speed improvement of ~450%. Figure 9(b) depicts the relationship between the trap closing time and the number of distinct annihilation events after the dump of the trapped antiprotons.

### C. System uptime and error handling

The TALOS framework has demonstrated a very high uptime, running since its first deployment in 2021 almost continuously (at least idle). The only moments offline were during system upgrades. A few reboots were needed during the first year due to unrecoverable error states. With the consolidation of the code, such reboots have become less and less necessary: only two were needed in 2023.

During the AE$\bar{g}$IS 2022 antiprotons campaign (spanning 35 days), the control system conducted measurements for 552.3 h (almost 22 days), equivalent to ~62% of the total time (which corresponds to most of the nights and weekends of the period. Daytime was mainly devoted to development). Throughout this period, the system faced various situations that prevented measurements due to external factors, like conditions identified by the *ELENA Interface*, differences in run conditions and data rates causing congestion, and variations in the time needed by the DAQ to sync data to disk within a fixed timeout, or other minor hardware or software-related exceptions, with *Retry* as the associated action.

In Table I, a summary of these exceptions is displayed, totaling ~66 h 27 min, which accounts for 12.7% of the measurements' total time and 7.9% of the entire antiproton campaign. This capability streamlines both data collection and data analysis. Without the ability to react to these exceptions, identifying and manually reacquiring affected runs would have imposed a significant overhead on the scientists.

In the summer of 2023, the AE$\bar{g}$IS collaboration performed its first experimental campaign on Highly Charged Ions (HCIs). Antiprotons were trapped as usual, and the interaction with nitrogen injected into the chamber was probed. During $\bar{p}$ collisional cooling with the gas, HCIs are formed either by collisional ionization or by antiproton's capture, which leads to a cascade of electrons emitted while it falls on the nucleus. In this experimental campaign (32 days), TALOS acquired data for 516.7 h (almost 22 days), equivalent to ~67% of the total time; by correctly handling the exceptions presented in the paragraph above, it saved ~16 h 21 min (i.e., 31.2% of the measurements' total time and ~21% of the entire campaign). The detailed results are presented in Table II.

### D. Safety

The ABORT system, running quietly in the background and typically unnoticed, once prevented potential hardware damage during an unsupervised nightly data collection. On that occasion, the high-voltage power supply connected to the Multi-Channel Plate (MCP) began to fail while biasing the MCP's front face at 2800 V. The corresponding μService reported a lost connection error, and TALOS raised ABORT (being a potentially hardware-damaging error, it was given *Criticality Code* 4, the highest). During the transition to *Safe Mode*, the system shut off all the high-voltage power supply lines. On the same night, a minor vacuum incident occurred, which could have caused damage to the MCP if it had remained continuously powered on.

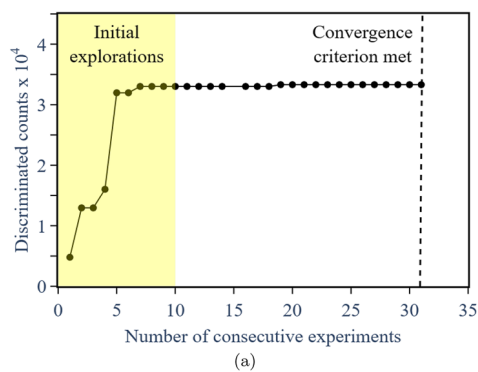 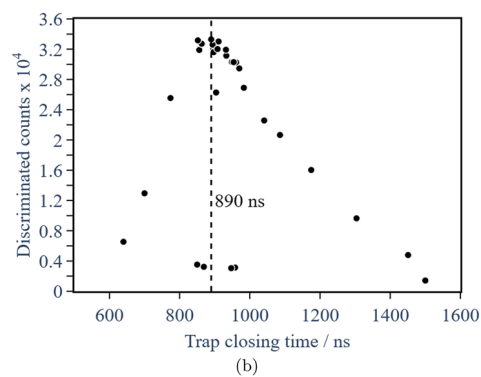

**FIG. 9.** Results of the trap closing time optimization: (a) the convergence plot of the highest observed number of annihilation events and (b) the number of observed annihilation events as a function of the trap closing time. The latter result is different from the one in 2022 (Fig. 6) because the capture electrode voltage was raised from 10 to 15 kV.

TABLE I. Details of the handling of the most frequent exceptions in the AEḡIS experiment during the antiproton campaign of 2022. With "Events," we mean the number of occurrences, while "Blocks" is the number of groups of contiguous scripts where the error keeps on appearing. "Total duration" is the total script time that was invalidated by the corresponding exception. The column "Total" is not simply the sum of all the previous columns: in fact, the exceptions that were thrown during the same script execution are cumulated and counted as one. This is especially necessary to correctly evaluate the total time.

|  | Beam stopper in | Valve closed | Empty shot | No triggers | DAQ | Total |
|---|---|---|---|---|---|---|
| Events | 330 | 172 | 27 | 251 | 864 | 1644 |
| Blocks | 66 | 28 | 19 | 48 | 148 | 244 |
| Total duration | 20 h 31′ | 13 h 3′ | 2 h 56′ | 27 h 39′ | 24 h 40′ | 66 h 27′ |

This event demonstrated the importance of having a distributed system capable of reacting to hardware exceptions in full autonomy, enhancing and ensuring the safety (and self-preservation) of the experiment.

### E. Synchronization

The AEḡIS experiment requires precise control of multiple subsystems to produce the antihydrogen. The manipulation of particles with electrodes, laser pulses, actuators, and other pieces of hardware needs to be operated with high synchronicity. The lifetime of positronium atoms puts a constraint of 10–100 ns for a single operation. Furthermore, the pulse length of the Ps-excitation laser further tightens the restriction to sub-10 ns precision. Because of that, the synchronicity between controllers responsible for the most sensitive operations is of the most importance. Moreover, to establish a continuous production of antihydrogen, it is required to have at least two Kaslis operating in parallel: one for controlling the 5 T trap for antiproton accumulation and the other for manipulating plasma in the 1 T region for antihydrogen production.

A series of dedicated measurements have been performed to measure the synchronization and delays in the system. First of all, the average round trip time for an echo message (a message sent in one direction, which gets re-sent back immediately after arrival) between two μServices running on different computers was measured to be 3(1) ms, while if the μServices are executed on the same PC, the time needed is in the order of tens of microseconds. These values give an estimation of the latency caused by the TALOS messaging system.

To test the performance of the parallel operations described in Sec. II F 2, a simple script was defined, consisting of a waiting routine, and a total of 5 Kaslis have been used, so to put the system under stress. For each parallel mode (asynchronous and synchronous), two series of 50 runs of the aforementioned script were executed. The first series was done with the same wait time of 2 s for all the 5 Kaslis used; in the second series, the wait time increased by 2 s between Kaslis, so from 0 s for the first Kasli to 8 s for the fifth Kasli.

To quantify the results, a measure, denoted as $\delta T$, was defined. To obtain it, at each iteration of the script, the five starting times of the script (one per Kasli) are taken and ordered. Then, four $\delta T$ are evaluated as the difference between each pair of contiguous elements (e.g., $T_2 - T_1$ and $T_3 - T_2$). This measure gives the overview of the expected synchronization coming from this software implementation.

Figures 10 and 11 show the calculated values of $\delta T$ at different series iterations as a box plot. Each point shows the range of all values, with the top and bottom of the box showing the 75 and 25 percentiles of measured values. Each plot is accompanied by the calculated average of $\delta T$, with error bars calculated as standard deviations.

Noticeable trends in asynchronous operation are evident, as the values of $\delta T$ tend to increase for the same-duration test [Fig. 10(a)] until reaching a plateau. This behavior can be attributed to the intrinsic delays included in Tamer and Monkey. The delays were added as a precaution to possible race conditions. These effects, however, are perfectly normal and demonstrate the asynchronicity of the operations.

TABLE II. Details of the handling of the most occurring exceptions in the AEḡIS experiment during the highly charged ions' campaign of 2023. The meaning of the various terms is the same as in Table I.

|  | Beam stopper in | Valve closed | Empty shot | No triggers | DAQ | Total |
|---|---|---|---|---|---|---|
| Events | 105 | 189 | 421 | 299 | 1601 | 2615 |
| Blocks | 42 | 39 | 83 | 48 | 129 | 242 |
| Total duration | 14 h 5′ | 27 h 26′ | 62 h 11′ | 50 h 37′ | 30 h 40′ | 161 h 21′ |

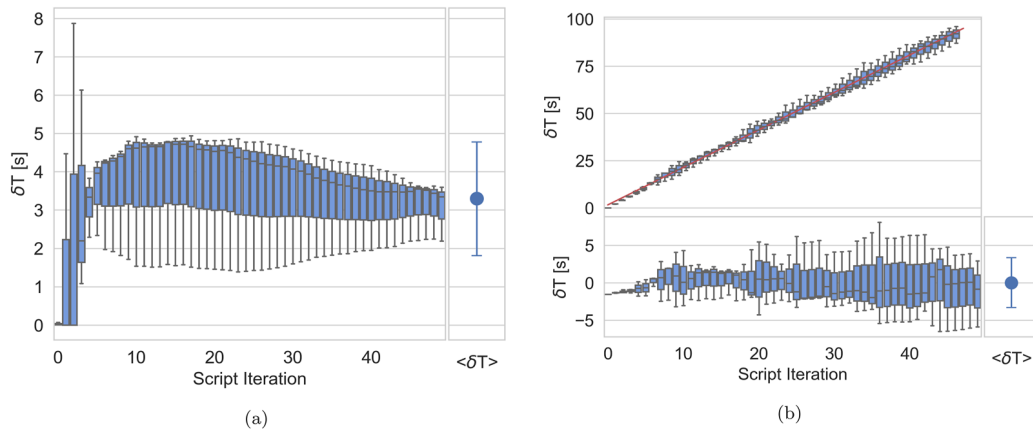

**FIG. 10.** Synchronization results in the asynchronous parallel operation mode, showing the time difference between the start of different parallel scripts, in a schedule of 50 scripts. Both plots are accompanied by a single point plot adjacent on the right, representing the average $\delta T$ over all 50 runs. In the same-duration script case (a), the high average can be explained by intrinsic delays included in Tamer and Monkey to avoid race conditions. In the different-duration script case (b), a linear trend is visible that comes from the increase in the duration time between different Kaslis. The bottom plot shows $\delta T$ corrected for the linear trend, showing the average jitter value of −5 to 5 s.

In the asynchronous operation, with different durations of the scripts [Fig. 10(b)], the trend is a linear increase with a slope of around 2 s per iteration. This is exactly what would be expected. In fact, the BANANA message always arrives later for Kaslis with a longer script duration. So, for the first Kasli, the end of the scripts (and therefore, the start of the subsequent one) happens approximately at 2, 4, 6 s, etc., while for the second one, it occurs approximately at 4, 8, 12 s, etc., hence the presence of the linear trend in $\delta T$.

On the contrary, when looking at the synchronous operation results, it is clear that there is no trend whatsoever. In both the cases of same-duration script [Fig. 11(a)] and different-duration script [Fig. 11(b)], the average $\delta T$ is below 10 ms. This measurement clearly shows the synchronization of the multiple scripts running in parallel and gives a precise value of the jitter to be expected between the start of different scripts on parallel Kaslis.

This analysis has proven the correct handling of (a) synchronicity by TALOS in the case of multi-Kasli operations, with a delay in launching synchronous scripts in the order of 10 ms. To enable the complex operations needed for antihydrogen production and study, a further level of synchronization has been inserted, at the script level, between Kaslis: using two digital lines between a pair of Kasli units, a low-level handshake was realized, creating a software barrier that ensures the synchronization to the nanosecond of the Python code following the barrier. The combination of the correct handling of parallel synchronous Kaslis by TALOS and the low-level barrier enables the various FPGA to operate synchronously with nanosecond precision, a requirement to efficiently form antihydrogen in the AEḡIS experiments (mainly given by the Ps excitation laser pulse length, as explained in the beginning of this subsection).

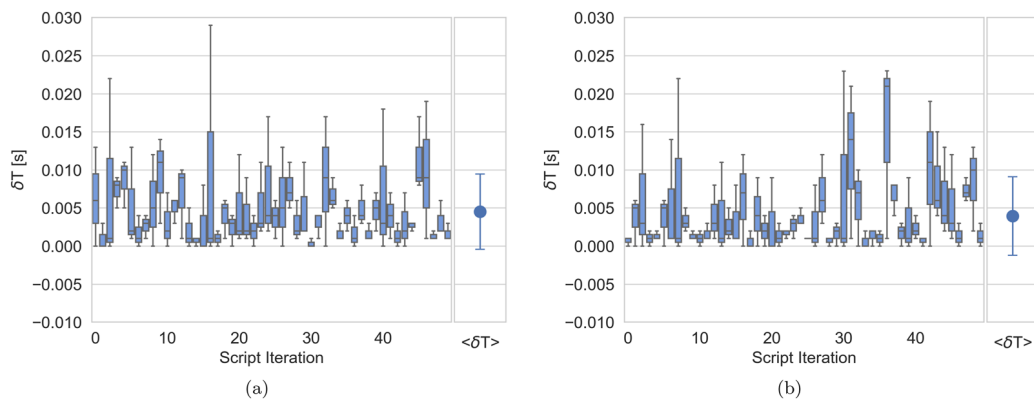

**FIG. 11.** Synchronization results in the synchronous parallel operation mode, showing the time difference between the start of different parallel scripts, in a schedule of 50 scripts. Both in the same-duration case (a) and in the different-duration case (b), the time difference between the start of different parallel scripts is stable around 5 ms, independently of the duration of the scripts. Both plots are accompanied by a single point plot adjacent on the right, representing the average $\delta T$ over all 50 runs.

It is worth commenting that the tests have been done using dummy Kaslis because more than two Kaslis were required in parallel for the test to have meaningful statistics, and only two were available at AEḡIS. However, the system was tested within the experiment without showing any downsides.

## V. CONCLUSION

In this article, the framework for autonomous control systems developed for the AEḡIS experiment, TALOS, is presented. It has been specifically created as the engine for a distributed control system, constituted of separate elements, called μServices, running in parallel asynchronously. At the same time, the entire computer architecture is masked, rendering effectively the entire experiment's network a single entity. The modularity of the system has proven crucial for continuous development and integration: every μService is coded and tested singularly and, therefore, can be brought to maturity before being inserted into the running system, maximizing coding efficiency and minimizing system downtime. This modality also ensures stability and extendability. In addition, the unification of the totality of the machines of the experiment into a single unit has enabled a very high degree of automation: in fact, high-level decisions, which are based on the simultaneous knowledge of multiple parameters originating from different points of the network, are possible and effective for this reason. In this regard, two design features have proven especially crucial: the communication layer, reliable and fast, and the error management scheme—from substitution to concentration, to the criticality system.

The automation is further enhanced by the capacity to make high-level decisions also based on the feedback of the data acquired. The culmination of this feature is autonomous parameter optimization, where TALOS can find the best parameter values that bring an observable as close as possible to the desired target.

Clearly, the application in the AEḡIS experiment was possible thanks to the deep integration with the Kasli, for the nanosecond-precision control of the time-sensitive routines.

All these characteristics have completely changed the way the scientists operate and interface with the experiment: from continuous shifts of operators constantly manipulating the control parameters of the experiment to ensure its correct behavior, to long periods of unsupervised data taking, where parameters are either automatically maintained to pre-defined setpoints, or autonomously adjusted to scan over opportune phase spaces or to optimize the values of observables. This has particularly proven vital with the introduction of ELENA, which delivers antiprotons to the experiment uninterruptedly (in the absence of faults). The automated system has maximized the amount of beam taken.

All this is proven by the results obtained in the past years, i.e., (i) the optimization, both via a full parameter space scanning, followed by manual analysis, and in total autonomy, of the best values of the ELENA beam position and the trap closure time; (ii) the reliable synchronization between multiple Kaslis, with a jitter of 10 ms; and (iii) the overall autonomy and error recovery, which resulted in the system running autonomously for an average of 64.6% of the experimental campaigns, saving more than 1000 hours by correctly handling exceptions, also externally generated. All these results lead to the record trapping of antiprotons in the AD: together with a more efficient $e^+ \to$ Ps converter and Ps excitation laser (also already demonstrated), this result will enable, in the next years, to produce antihydrogen with a rate of at least 1 $\bar{H}$/min, which is 2–3 orders of magnitude higher with respect to what has been achieved previously.

Moreover, it allows the exploration of new physics avenues, by enabling the jitter-correction of the laser that rendered possible the positronium laser cooling,[27] and the continued model optimization that led to the formation of highly charged ions in the AEḡIS Penning–Malmberg trap.

All these results demonstrate that the TALOS framework is a powerful tool for building control systems that are reliable, extendable, and maintainable and that guarantee reproducible scientific results. Important to notice is that the framework is experiment-agnostic and possesses the potential for applications beyond the realm of antimatter, quantum, and atomic physics, potentially serving a wider spectrum of scientific domains. Notably, this framework is released as open-source (DOI: 10.5281/zenodo.10371404), allowing other experiments to benefit from the collaborative endeavor.

## ACKNOWLEDGMENTS

The AEḡIS collaboration acknowledges the following funding agencies for their support.

This work has been financed by the CERN Doctoral Student Program and the Istituto Nazionale di Fisica Nucleare (INFN)—Sezione di Trento.

This work has been sponsored by the Wolfgang Gentner Program of the German Federal Ministry of Education and Research (Grant No. 13E18CHA).

This work was funded by the Research University—Excellence Initiative of Warsaw University of Technology via the Young PW program under Agreement No. CPR-IDUB/56/Z01/2024 and via the strategic funds of the Priority Research Center of High Energy Physics and Experimental Techniques by the Polish National Science Center under Agreement Nos. 2022/45/B/ST2/02029 and 2022/46/E/ST2/00255, and by the Polish Ministry of Education and Science under Agreement No. 2022/WK/06.

The AEḡIS collaboration thanks the CERN accelerator and decelerator teams for the outstanding performance of the AD–ELENA complex.

We thank Odd Oyvind Andreassen and the rest of the CERN LabVIEW Support team for the illuminating discussions and the code reviews.

We also thank Dr. Simone Stracka for the inspiring discussions.

## AUTHOR DECLARATIONS
### Conflict of Interest

The authors have no conflicts to disclose.

### Author Contributions

**M. Volponi**: Conceptualization (supporting); Data curation (equal); Formal analysis (equal); Software (lead); Supervision (equal); Writing – original draft (lead); Writing – review & editing (lead). **J. Zieliński**: Data curation (supporting); Formal analysis (equal);

Software (equal); Writing – original draft (supporting). **T. Rauschendorfer**: Formal analysis (equal); Software (supporting); Writing – original draft (supporting). **S. Huck**: Data curation (equal); Software (supporting). **R. Caravita**: Conceptualization (lead); Data curation (equal); Formal analysis (equal); Supervision (equal). **M. Auzins**: Writing – review & editing (supporting). **B. Bergmann**: Writing – review & editing (supporting). **P. Burian**: Writing – review & editing (supporting). **R. S. Brusa**: Writing – review & editing (supporting). **A. Camper**: Writing – review & editing (supporting). **F. Castelli**: Writing – review & editing (supporting). **G. Cerchiari**: Writing – review & editing (supporting). **R. Ciuryło**: Writing – review & editing (supporting). **G. Consolati**: Writing – review & editing (supporting). **M. Doser**: Writing – review & editing (supporting). **K. Eliaszuk**: Writing – review & editing (supporting). **A. Giszczak**: Writing – review & editing (supporting). **L. T. Glöggler**: Writing – review & editing (supporting). **Ł. Graczykowski**: Writing – review & editing (supporting). **M. Grosbart**: Writing – review & editing (supporting). **F. Guatieri**: Writing – review & editing (supporting). **N. Gusakova**: Writing – review & editing (supporting). **F. Gustafsson**: Writing – review & editing (supporting). **S. Haider**: Writing – review & editing (supporting). **M. A. Janik**: Writing – review & editing (supporting). **T. Januszek**: Writing – review & editing (supporting). **G. Kasprowicz**: Writing – review & editing (supporting). **G. Khatri**: Writing – review & editing (supporting). **Ł. Kłosowski**: Writing – review & editing (supporting). **G. Kornakov**: Writing – review & editing (supporting). **V. Krumins**: Writing – review & editing (supporting). **L. Lappo**: Writing – review & editing (supporting). **A. Linek**: Writing – review & editing (supporting). **J. Malamant**: Writing – review & editing (supporting). **S. Mariazzi**: Writing – review & editing (supporting). **L. Penasa**: Writing – review & editing (supporting). **V. Petracek**: Writing – review & editing (supporting). **M. Piwiński**: Writing – review & editing (supporting). **S. Pospisil**: Writing – review & editing (supporting). **L. Povolo**: Writing – review & editing (supporting). **F. Prelz**: Writing – review & editing (equal). **S. A. Rangwala**: Writing – review & editing (supporting). **B. S. Rawat**: Writing – review & editing (supporting). **B. Rienäcker**: Writing – review & editing (supporting). **V. Rodin**: Writing – review & editing (supporting). **O. M. Røhne**: Writing – review & editing (supporting). **H. Sandaker**: Writing – review & editing (supporting). **P. Smolyanskiy**: Writing – review & editing (supporting). **T. Sowiński**: Writing – review & editing (equal). **D. Tefelski**: Writing – review & editing (supporting). **T. Vafeiadis**: Writing – review & editing (supporting). **C. P. Welsch**: Writing – review & editing (supporting). **T. Wolz**: Writing – review & editing (supporting). **M. Zawada**: Writing – review & editing (supporting). **N. Zurlo**: Writing – review & editing (supporting).

## DATA AVAILABILITY

The data supporting the findings of this study are available under the conditions of the AEḡIS Collaboration Data Management Plan.

## APPENDIX: TALOS IMPLEMENTATION DETAILS

### 1. The Guardian

The Guardian is not simply the *Root Actor* of all μServices on each machine, but is the real core of TALOS: in fact, it provides the messaging system for the μServices so that they can interact as if they were on the same computer, and it maintains the three watchdogs that guarantee the reliability of the system. They consist of the following:

- Guardians watchdog: Each Guardian sends periodically a special message to every other Guardians; the times of arrival of the last received messages are checked, and if too old, the system is put in *Safe Mode*. The *Safe Mode* is a special state of the system where everything is put to idle, to minimize eventual damage caused by, for example, a hardware fault. Manual intervention is needed to bring the system out of this state;
- μServices watchdog: Similarly to the Guardians watchdog, each Guardian checks the messages periodically sent by the μServices running on the same machine. In case of unresponsiveness, the Guardian tries to restart the frozen μService before halting system operations;
- `ABORT` watchdog: This process keeps the `ABORT` *Shared Variable* (SV) unique on the entire experiment network. The `ABORT` SV is used to effectively propagate the necessity of putting the system in *Safe Mode* among the various Guardians. This watchdog periodically scans the network to check the SV's accessibility. If not detected, a new local instance is deployed and activated. In cases of multiple instances found (e.g., due to temporary network disconnections), the copy whose name comes later in alphabetical order is undeployed. While maintaining `ABORT` as a network-wide singleton is challenging, this approach ensures the presence of at least one copy, and its value is unequivocal.

The driving requirement for these watchdogs was reliability over speed: not to overload the system with messages, the characteristic replies that testing time is in the order of seconds. The typical communication speed among Guardians (millisecond range) is negligible for the purpose of the watchdogs. Each Guardian does not assume that all PCs on the network are part of the experiment: the network boundary is set out in the shared configuration file.

### 2. μService structure and the FOAM

All the μServices are children of FOAM; therefore, they are all *Nested Actors* of the Guardian. This inheritance structure has several advantages: (i) it masks completely the complexity of the Actor Framework and of the rest of the internal mechanics of TALOS, to the end-user/developer; (ii) it gives the possibility of updates to the framework without affecting the per-μService specific code; and (iii) it automates and/or enables several functions via dedicated methods.

By inheriting from FOAM, all the μServices become structured as a *Queued Message Handler* (QMH), i.e., the combination of a part reacting to external or internal events, generally referred to as *Event Handler*, and a *Consumer*, which executes the queue of tasks populated both by preceding tasks and by the event handler part.

To host the μService-specific code, each μService has to override a certain number of VIs, each serving a specific function, as follows:

- < μService Name > ctl: This is the μService actor/class private data container.

- Init: This VI is executed before the μService is started: it is useful to perform initial checks and to populate the values of the private data of the μService.
- Close: This is used to safely close everything opened in *Init* and during the μService execution. This VI is guaranteed to be executed during the μService shutdown procedure.
- Consumer: This VI executes the messages in the internal queue of the μService, dequeuing them in the order they are sent and according to their priority.
- Event handler: Every time an event related to the μService is triggered, this VI is called. The event types available are button pressed, shared variable updated, message (e.g., from another μService), ready to launch? (called at the beginning of each schedule to assess each μService health), start run, stop run, and safe mode.
- μService GUI: The front panel of this VI is the actual GUI of the μService.
- Process error: This VI gets called every time an error is generated in whatever part of the μService. This VI should serve as a first filter of the errors; otherwise, the error is forwarded to the distributed system *Error Manager*.

The execution of multiple tasks is eased by several methods, which every μService inherits from FOAM. They are available as a dedicated palette that can be installed in LabVIEW, to be quickly accessible during the coding of any μService. Their functionalities span from getting or updating a value from the GUI to sending a message to another μService, enqueuing a new (possibly periodic) task to the *Consumer*, generating (or substituting) a custom-defined error, launching another μService, and many others.

#### 3. The Error Manager

A key part of the automation provided by TALOS lies in the distributed error management system. In particular, the fact that the system will respond to every error generated in any of its parts is the core foundation of the safety of TALOS. The three key concepts upon which the error management system is based are the following:

- **Error substitution** relies on converting general error codes into user-defined ones, which also carry context-specific information: in fact, the same error can have a completely different meaning (and so require a different action to be taken) if generated in different contexts.
- **Error concentration**, instead, is the process of gathering all the errors generated by the various parts of the distributed system to a single μService, the *Error Manager*. This solution ensures the generation of a consistent response to every error, regardless of its source.
- **Error criticality** is defined by assigning a numerical value to every error, ranging from 0 to 4, indicating the appropriate action (continue, retry, skip, stop, and abort) to be taken in response to that error. The system keeps track of the criticality codes generated during the experiment execution and, upon completion, the action corresponding to the highest criticality level encountered is performed (see Sec. II E).

The strength of this error management system lies in its simplicity: in fact, the three key concepts have proven to guarantee its reliability, stability, and scalability to very high numbers of user-defined errors (more than 500, in AEḡIS at the moment of writing). Conversely, it is not "too simple": the five levels of error criticality give to the system the ability to respond flexibly in every situation.

#### 4. The Scheduler

An indispensable functionality of any autonomous system is the ability to define a sequence of tasks the system needs to perform. For this reason, the Scheduler μService was created to allow users to define a schedule of scripts the system will execute.

A schedule is defined in Schedule Blocks (SBlocks), each defining a series of runs that execute the same script with possibly varying parameters. There are two types of SBlocks: *Scan* SBlocks perform a pre-defined scan over some script parameter lists, while *Optimization* SBlocks use the integration of ALPACA (see Sec. II D) to autonomously explore the parameter space to find optimal values of some parameters to bring some pre-defined observable as close as possible to a target value. To add new SBlocks, the user can use the *Add Run* window as shown in Fig. 12. The script to be executed can be chosen via a browser, and the *Common Parameters*, fixed throughout the entire SBlock, can be defined. Furthermore, the *Quality of Run* can be specified, to retry the script if this quality criterion is not met (see Sec. II F 4).

If the scan mode is selected [Fig. 12(a)], the sequence of parameters to scan over has to be defined: either the full list can be given, or (if numeric) the *start*, *stop*, and *step* (size) values can be provided, and the *Scheduler* will generate the corresponding list in linear, exponential, or linked progression. When two (or more) parameters are linked together, the scan is performed varying them together, using the *No. of steps* value [from the Add Run window shown in Fig. 12(a)] to calculate the step size.

In the optimization mode [Fig. 12(b)], the user defines the parameters (numbers and/or strings) that are used to perform the optimization over, together with their boundaries, and the list of observables that the optimizer uses to evaluate the run results, together with their optimization strategy, i.e., maximum, minimum, or specific value.

All created SBlocks can be modified, and their order can be changed at any time when creating schedules. The Scheduler supports the creation of multiple schedules when working in the parallel operation mode (see Sec. II F 2 for more details).

#### 5. The Monkey

The *Monkey* plays a central role in TALOS automation, taking high-level decisions that typically fall through to the user. It executes the sequence of scripts defined in the Scheduler (see Sec. II E), and it determines the action to be taken upon each script completion based on a value (ranging from 0 to 4) that is evaluated as the highest value among the error criticalities and `BANANA`, where the error criticalities are the ones accumulated during the script run, and the `BANANA` is the value returned from Kasli based on the script execution. Depending on the evaluated value, the possible actions are as follows:

- 0—Everything finished well. The system will execute the next script.

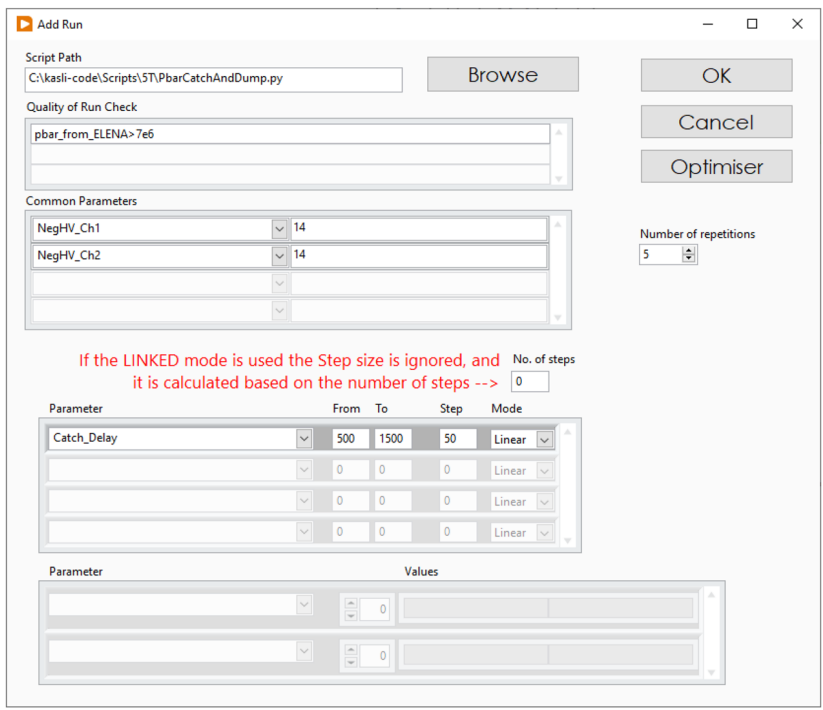

(a)

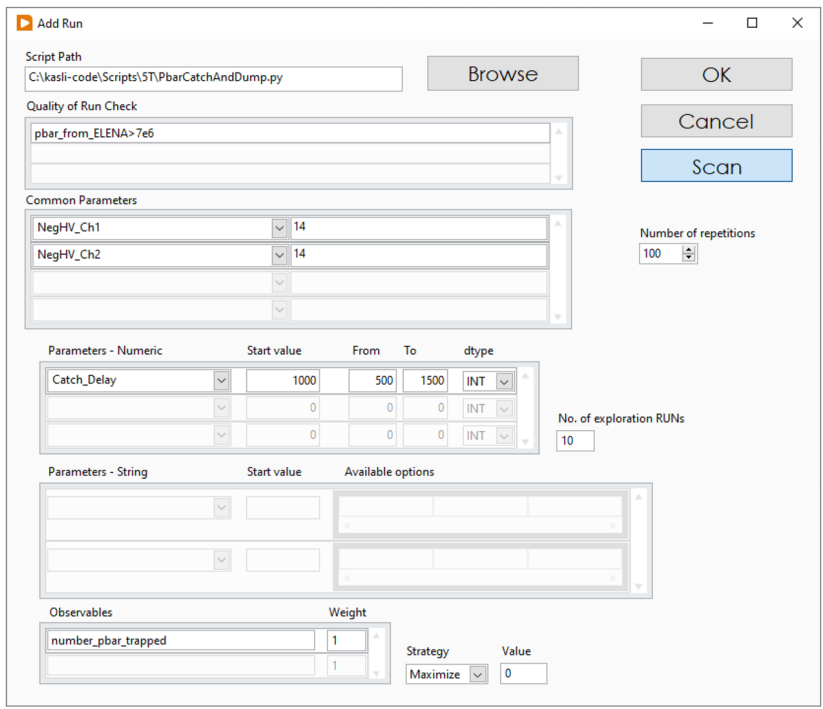

(b)

**FIG. 12.** *Add Run* panels for different schedule blocks. The selector for the script choice and the first table for the fixed parameters of the execution are always present. In the scan case (a), multiple tables are available to define the parameter sequences to scan over. In the optimization case (b) instead, the tables defining the parameters and the ranges over which they can be optimized by evaluating the observable(s) are edited at the bottom of the window.

- 1—There was a minor problem in the execution (e.g., some data were not saved): the script will be re-executed to ensure that the given configuration of parameters is correctly measured.
- 2—A problem prevented the completion of the script, e.g., an incorrect configuration was provided. Since retrying the execution will give the same result, the script will be skipped, and the system will move on to the subsequent one.
- 3—There was an error during the operation, e.g., the Kasli controller not responding. In this case, the entire schedule will be skipped and the system will remain idle.
- 4—A critical error happened: the *Monkey* skips the entire schedule and stops the execution of the current script. Meanwhile, the Guardians will put the entire control system in *Safe Mode*.

All the skipped scripts are saved in a new schedule called *Skipped RUNs*. They can later be sent back to the Scheduler for editing, or run again.

Upon the correct ending of a script from an Optimization SBlock, the *Monkey* contacts ALPACA to retrieve the new parameter values to be used on the subsequent re-execution of the script (or end the optimization if the target is reached).

## 6. Detector Manager and Father of All Detectors

The AE$\bar{g}$IS experiment, like most physics experiments, is characterized by a high number of detectors. Albeit very different in nature and function, their pattern of operation essentially boils down to (1) the configuration for the acquisition of the signal, (2) the acquisition phase itself, and (3) the saving of captured data.

In light of this general schematization, the combination of a μService, *Detector Manager*, and a hardware class, *Father Of All Detectors* (FOAD), was created so that every time a new μService needs to be generated to manage a new detector, it is only necessary to create two children, one per each of the classes just mentioned, and fill their components.

The class FOAD represents the generalization of the detector hardware functionality: each of its VIs (*Init*, *Set Config*, *Arm*, *Acquiring*, *Save Data*, *Stop*, and *Close*) represents a specific action for the detector. Each child, which implements the software interface with a real detector, needs to override them.

*Detector Manager* contains all the instructions needed for the correct functioning of the μService, from the message interaction with the rest of the system to the GUI. In each of its children, it is only needed to specify the corresponding child of FOAD to manage. Together with some flags, namely *Auto-ReArm*, *Stop after Save Data*, and *Stop before ReArm*, multiple patterns can then be obtained, which cover most cases of detectors to implement.

## 7. The CIRCUS graphical user interface

An example of the Graphical User Interface (GUI) of the CIRCUS control system, provided by TALOS, is shown in Fig. 13. In the upper left corner, the displays of the Guardians and μServices watchdog statuses are located, while on the upper right, a list containing the most recent errors is present. In the column on the right, the details of an error can be visualized by simply selecting it, and on the bottom, the real-time log indicates the Kasli(s) current activity. This part of the GUI is identical (and displays exactly the same information) on all the computers in the system. Inside the μService subpanel, the *Tamer* is displayed, in the act of managing two *Monkeys* each executing a schedule on the corresponding Kasli.

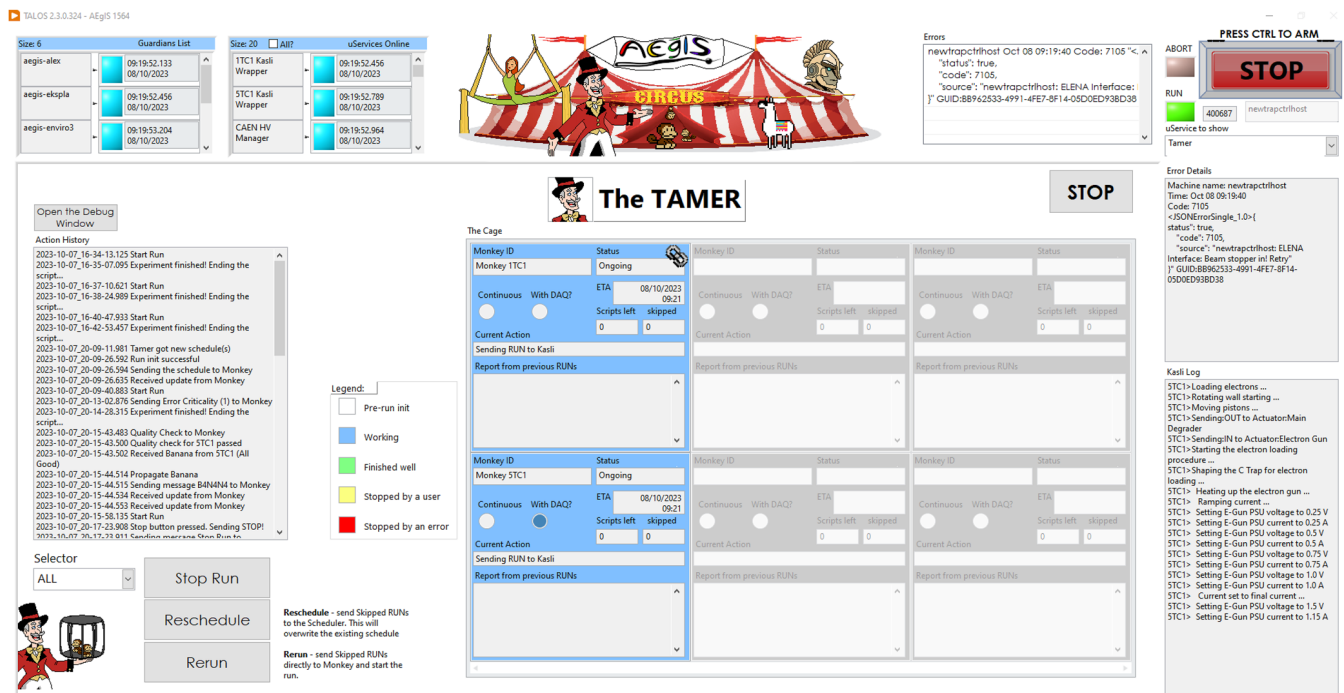

**FIG. 13.** A screenshot of the CIRCUS control system executing a schedule of experiments involving antiprotons. The primary interface is provided by TALOS. Located in the upper left corner are the Guardians and μServices watchdogs, while the error list is positioned in the upper right corner. In the right-hand column, the top section provides specific details regarding the selected error and, underneath it, a real-time log displays Kaslis operational activities. This interface is uniform across all experiment machines. Within the main window, the *Tamer* μService is displayed, presently overseeing two *Monkeys* performing an active measurement schedule with two Kaslis.

## REFERENCES

[1]M. Doser, S. Aghion, C. Amsler, G. Bonomi, R. S. Brusa, M. Caccia, R. Caravita, F. Castelli, G. Cerchiari, D. Comparat, G. Consolati, A. Demetrio, L. Di Noto, C. Evans, M. Fanì, R. Ferragut, J. Fesel, A. Fontana, S. Gerber, M. Giammarchi, A. Gligorova, F. Guatieri, S. Haider, A. Hinterberger, H. Holmestad, A. Kellerbauer, O. Khalidova, D. Krasnický, V. Lagomarsino, P. Lansonneur, P. Lebrun, C. Malbrunot, S. Mariazzi, J. Marton, V. Matveev, Z. Mazzotta, S. R. Müller, G. Nebbia, P. Nedelec, M. Oberthaler, N. Pacifico, D. Pagano, L. Penasa, V. Petracek, F. Prelz, M. Prevedelli, B. Rienaecker, J. Robert, O. M. Røhne, A. Rotondi, H. Sandaker, R. Santoro, L. Smestad, F. Sorrentino, G. Testera, I. C. Tietje, E. Widmann, P. Yzombard, C. Zimmer, J. Zmeskal, and N. Zurlo, "AEgIS at ELENA: Outlook for physics with a pulsed cold antihydrogen beam," Philos. Trans. R. Soc., A 376, 20170274 (2018).

[2]C. Amsler, M. Antonello, A. Belov, G. Bonomi, R. S. Brusa, M. Caccia, A. Camper, R. Caravita, F. Castelli, P. Cheinet, D. Comparat, G. Consolati, A. Demetrio, L. Di Noto, M. Doser, M. Faní, R. Ferragut, J. Fesel, S. Gerber, M. Giammarchi, A. Gligorova, L. T. Gl oggler, F. Guatieri, S. Haider, A. Hinterberger, A. Kellerbauer, O. Khalidova, D. Krasnický, V. Lagomarsino, C. Malbrunot, S. Mariazzi, V. Matveev, S. Müller, G. Nebbia, P. Nedelec, L. Nowak, M. Oberthaler, E. Oswald, D. Pagano, L. Penasa, V. Petracek, L. Povolo, F. Prelz, M. Prevedelli, B. Rienäcker, O. Rohne, A. Rotondi, H. Sandaker, R. Santoro, G. Testera, I. Tietje, V. Toso, T. Wolz, P. Yzombard, C. Zimmer, and N. Zurlo, "Pulsed production of antihydrogen," Commun. Phys. 4, 19 (2021).

[3]E. Perego, M. Pomponio, A. Detti, L. Duca, C. Sias, and C. E. Calosso, "A scalable hardware and software control apparatus for experiments with hybrid quantum systems," Rev. Sci. Instrum. 89, 113116 (2018).

[4]P. T. Starkey, C. J. Billington, S. P. Johnstone, M. Jasperse, K. Helmerson, L. D. Turner, and R. P. Anderson, "A scripted control system for autonomous hardware-timed experiments," Rev. Sci. Instrum. 84, 085111 (2013).

[5]J. Agraz, A. Grunfeld, D. Li, K. Cunningham, C. Willey, R. Pozos, and S. Wagner, "LabVIEW-based control software for para-hydrogen induced polarization instrumentation," Rev. Sci. Instrum. 85, 044705 (2014).

[6]A. Trenkwalder, M. Zaccanti, and N. Poli, "A flexible system-on-a-chip control hardware for atomic, molecular, and optical physics experiments," Rev. Sci. Instrum. 92, 105103 (2021).

[7]A. Keshet and W. Ketterle, "A distributed, graphical user interface based, computer control system for atomic physics experiments," Rev. Sci. Instrum. 84, 015105 (2013).

[8] As an example, the AE̅gIS experiment is undergoing an extension in order to be able to explore the physics of antiprotonic atomic systems, which was not foreseen when the experiment was conceived. Conversely, big observational experiments are less inclined to extensive modifications.

[9]M. Volponi, S. Huck, R. Caravita, J. Zielinski, G. Kornakov, G. Kasprowicz, D. Nowicka, T. Rauschendorfer, B. Rienäcker, F. Prelz, M. Auzins, B. Bergmann, P. Burian, R. S. Brusa, A. Camper, F. Castelli, G. Cerchiari, R. Ciurylo, G. Consolati, M. Doser, K. Eliaszuk, A. Giszczak, L. T. Glöggler, L. Graczykowski, M. Grosbart, F. Guatieri, N. Gusakova, F. Gustafsson, S. Haider, M. A. Janik, T. Januszek, G. Khatri, L. Klosowski, V. Krumins, L. Lappo, A. Linek, J. Malamant, S. Mariazzi, L. Nowak, E. Oswald, L. Penasa, V. Petracek, M. Piwinski, S. Pospisil, L. Povolo, S. A. Rangwala, B. S. Rawat, V. Rodin, O. M. Røhne, H. Sandaker, P. Smolyanskiy, T. Sowinski, D. Tefelski, C. P. Welsch, T. Wolz, M. Zawada, and N. Zurlo, "Circus: An autonomous control system for antimatter atomic and quantum physics experiments," EPJ Quantum Technol. 11, 10 (2024).

[10]C. Hewitt, P. Bishop, and R. Steiger, "A universal modular ACTOR formalism for artificial intelligence," in *Proc. International Joint Conference on Artificial Intelligence* (Morgan Kaufmann Publishers, Inc., 1973), pp. 235–245.

[11]R. Bitter, T. Mohiuddin, and M. Nawrocki, *Labview: Advanced Programming Techniques*, 2nd ed. (Crc Press, Taylor and Francis Group, 2017).

[12]V. Cerf and R. Kahn, "A protocol for packet network intercommunication," IEEE Trans. Commun. 22, 637–648 (1974).

[13]G. Kasprowicz, P. Kulik, M. Gaska, T. Przywozki, K. Pozniak, J. Jarosinski, J. W. Britton, T. Harty, C. Balance, W. Zhang, D. Nadlinger, D. Slichter, D. Allcock, S. Bourdeauducq, R. Jördens, and K. Pozniak, "ARTIQ and sinara: Open software and hardware stacks for quantum physics," in *OSA Quantum 2.0 Conference*, edited by M. Raymer, C. Monroe, and R. Holzwarth (Optica Publishing Group, 2020), paper QTu8B.14.

[14]C. K. Lam, S. Maka, D. Nadlinger, C. Ballance, and S. Bourdeauducq, "Combining processing throughput, low latency and timing accuracy in experiment control," arXiv:2111.15290 [physics.ins-det] (2021).

[15]S. Bourdeauducq, R. Jördens, P. Zotov, J. Britton, D. Slichter, D. Leibrandt, D. Allcock, A. Hankin, F. Kermarrec, Y. Sionneau, R. Srinivas, T. R. Tan, and J. Bohnet, Artiq 1.0 (2016).

[16] At least for a person not completely new to Python: proficiency in the language is not needed for not-too-complex procedures.

[17]F. Pedregosa, G. Varoquaux, A. Gramfort, V. Michel, B. Thirion, O. Grisel, M. Blondel, P. Prettenhofer, R. Weiss, V. Dubourg, J. Vanderplas, A. Passos, D. Cournapeau, M. Brucher, M. Perrot, and E. Duchesnay, "Scikit-learn: Machine learning in Python," J. Mach. Learn. Res. 12, 2825–2830 (2011).

[18] It is customary to refer to multiple instances of the same μServices service by simply using the plural.

[19] Actually, its generality would make it suitable to handle also distributed systems that are not strictly experiments.

[20]M. Fischler, J. Lykken, and T. Roberts, "Direct observation limits on antimatter gravitation," arXiv:0808.3929 [hep-th] (2008).

[21]M. J. Borchert, J. A. Devlin, S. R. Erlewein, M. Fleck, J. A. Harrington, T. Higuchi, B. M. Latacz, F. Voelksen, E. J. Wursten, F. Abbass, M. A. Bohman, A. H. Mooser, D. Popper, M. Wiesinger, C. Will, K. Blaum, Y. Matsuda, C. Ospelkaus, W. Quint, J. Walz, Y. Yamazaki, C. Smorra, and S. Ulmer, "A 16-parts-per-trillion measurement of the antiproton-to-proton charge-mass ratio," Nature 601, 53–57 (2022).

[22]E. K. Anderson, C. J. Baker, W. Bertsche, N. M. Bhatt, G. Bonomi, A. Capra, I. Carli, C. L. Cesar, M. Charlton, A. Christensen, R. Collister, A. Cridland Mathad, D. Duque Quiceno, S. Eriksson, A. Evans, N. Evetts, S. Fabbri, J. Fajans, A. Ferwerda, T. Friesen, M. C. Fujiwara, D. R. Gill, L. M. Golino, M. B. Gomes Goncalves, P. Grandemange, P. Granum, J. S. Hangst, M. E. Hayden, D. Hodgkinson, E. D. Hunter, C. A. Isaac, A. J. U. Jimenez, M. A. Johnson, J. M. Jones, S. A. Jones, S. Jonsell, A. Khramov, N. Madsen, L. Martin, N. Massacret, D. Maxwell, J. T. K. McKenna, S. Menary, T. Momose, M. Mostamand, P. S. Mullan, J. Nauta, K. Olchanski, A. N. Oliveira, J. Peszka, A. Powell, C. Rasmussen, F. Robicheaux, R. L. Sacramento, M. Sameed, E. Sarid, J. Schoonwater, D. M. Silveira, J. Singh, G. Smith, C. So, S. Stracka, G. Stutter, T. D. Tharp, K. A. Thompson, R. I. Thompson, E. Thorpe-Woods, C. Torkzaban, M. Urioni, P. Woosaree, and J. S. Wurtele, "Observation of the effect of gravity on the motion of antimatter," Nature 621(7980), 716–722 (2023).

[23]M. Volponi, "Progress towards measuring the fall of antimatter in Earth's gravitational field," paper presented at the 57th Rencontres de Moriond on Gravitation (Moriond Gravitation 2023), La Thuile, Italy, 18-25 March 2023; available at https://inspirehep.net/literature/2666878.

[24]R. Caravita, "Progress report on the AEgIS experiment (2023)," CERN-SPSC-2024-002; SPSC-SR-339, 2023.

[25]S. Huck *et al.*, "Toward a pulsed antihydrogen beam for WEP tests in AEgIS," EPJ Web Conf. 282, 01005 (2023).

[26]W. Bartmann, P. Belochitskii, H. Breuker, F. Butin, C. Carli, T. Eriksson, S. Maury, R. Kersevan, S. Pasinelli, G. Tranquille, and G. Vanbavinckhove, "Progress in ELENA design," in *Proceedings of IPAC 2013, Shanghai (China)* (JACoW, 2013), pp. 2651–2653; available at https://accelconf.web.cern.ch/ipac2013/papers/wepea062.pdf.

[27]L. T. Gloggler, N. Gusakova, B. Rienacker, A. Camper, R. Caravita, S. Huck, M. Volponi, T. Wolz, L. Penasa, V. Krumins, F. Gustafsson, M. Auzins, B. Bergmann,

P. Burian, R. S. Brusa, F. Castelli, R. Ciurylo, D. Comparat, G. Consolati, M. Doser, L. Graczykowski, M. Grosbart, F. Guatieri, S. Haider, M. A. Janik, G. Kasprowicz, G. Khatri, L. Klosowski, G. Kornakov, L. Lappo, A. Linek, J. Malamant, S. Mariazzi, V. Petracek, M. Piwinski, S. Pospisil, L. Povolo, F. Prelz, S. A. Rangwala, T. Rauschendorfer, B. S. Rawat, V. Rodin, O. M. Rohne, H. Sandaker, P. Smolyanskiy, T. Sowinski, D. Tefelski, T. Vafeiadis, C. P. Welsch, M. Zawada, J. Zielinski, and N. Zurlo, "Positronium laser cooling via the $1^3s$–$2^3p$ transition with a broadband laser pulse," Phys. Rev. Lett. **132**, 083402 (2024).